\documentclass[preprintnumbers, prd, onecolumn, floatfix,amsmath,amsmath,
preprintnumbers,showpacs, letterpaper, superscriptaddress,nofootinbib]{revtex4}
\usepackage{graphicx,epsfig}
\usepackage{amsmath}
\usepackage {amssymb}
\usepackage{subfigure}
\usepackage{bm}
\usepackage{amssymb}
\usepackage{float}
\usepackage{amsmath}
\usepackage{subfigure}
\usepackage{braket}
\usepackage[colorlinks]{hyperref}
\usepackage[usenames,dvipsnames]{color}
\hypersetup{
	breaklinks=true,
	pdfstartview={FitH},    % fits the width of the page to the window
	colorlinks=true,       % false: boxed links; true: colored links
	linkcolor=blue,          % color of internal links
	citecolor=red,        % color of links to bibliography
	filecolor=magenta,      % color of file links
	urlcolor=blue,           % color of external links
	anchorcolor=green,      % Color for anchor text
	linktocpage=true
}

\def\doi{http://doi.org}

\def\r{\mathrm{r}}

\begin{document}
\newcommand\be{\begin{equation}}
\newcommand\ee{\end{equation}}
\newcommand\bea{\begin{eqnarray}}
\newcommand\eea{\end{eqnarray}}
\newcommand\bseq{\begin{subequations}} %solo con amsmath
\newcommand\eseq{\end{subequations}}
\newcommand\bcas{\begin{cases}}
\newcommand\ecas{\end{cases}}
\newcommand{\p}{\partial}
\newcommand{\f}{\frac}

\title{Final state of instabilities in Born-Infeld black holes    }
%Einstein-conformally Maxwell invariant gravity}

\author {A. Rahmani}\email{a$_$rahmani@sbu.ac.ir}
\
\affiliation {Department of Physics, Shahid Beheshti University, Evin, Tehran, Iran}

\author {M. Honardoost}\email{m$_$honardoost@sbu.ac.ir}
\affiliation{Department of Physics, Shahid Beheshti University,  Evin, Tehran, Iran}

\author {H. R. Sepangi}\email{hr-sepangi@sbu.ac.ir }
\affiliation{Department of Physics, Shahid Beheshti University,  Evin, Tehran, Iran}

\date{\today}
 %---------------------------------------

\begin{abstract}
We study two types of linear instabilities in Einstein-Born-Infeld-scalar field theory and show that small Born-Infeld (BI) black holes suffer from superradiant instability within a particular range of frequencies; the larger the BI coupling parameter, the slower the growth rate of the instability. It is predicted that a small BI black hole decays into a small hairy black hole that has a charged scalar condensate floating near the horizon. We numerically show that there is a phase transition between hairy and BI black holes. The metric solution shows the existence of a hairy black hole for $T<T_{c}$ and a BI black hole for $T>T_{c}$. In addition, the BI black hole will suffer from tachyonic mass, resulting in a near horizon scalar condensation instability, which is related to a planar hairy black hole. We show that the BI coupling parameter has a direct relation to the critical temperature in the small hairy black hole framework. This is not true for planar hairy black holes where a larger BI coupling parameter leads to a smaller critical temperature. Finally, it is shown that a small hairy black hole is stable at the critical temperature so it can be considered as the end point of superradiant instability.

\end{abstract}

\maketitle
\section{Introduction}
One of the interesting issues in black hole physics is the energy extraction from such objects which can occur in Kerr or charged black holes through scattering at the event horizon. This phenomenon is a generalization of the Penrose process and is known as superradiance \cite{pani, brito}. This has become a more interesting phenomenon since the existence of black holes has been observationally established in the past few years. The first direct verification of their existence was the detection of the gravitational wave signal GW150914  by LIGO, arisen from the collision and merger of a pair of black holes \cite{abo} and also the recent discovery of a supermassive black hole
at the core of the distant galaxy Messier $87^*$ by the Event Horizon Telescope
\cite{ref:eht}. In a superradiance scenario, a scalar field incident upon a charged or Kerr black hole scatters off with an enhanced amplitude in a certain frequency range. If there is a reflective boundary, for example a mirror \cite{pani,win} or an AdS boundary \cite{ads1,ads2,ads3,ads4,oscar,oscar1} or mass of the scalar field for the Kerr black hole \cite{masswin,massherd,mass2herd,hod}, under certain conditions the scattered wave bounces back and forth which would lead to exponential growth and instability. An important  problem is the final state of superradiant instability which can lead to the violation of no hair theorem \cite{rad,vol}. It is well known that the final state of superradiant instability could be a hairy black hole \cite{masswin,bosch} or a bosenova, namely an explosive event resulting from a full nonlinear investigation \cite{yosh,deg}. Such instabilities have also been shown to constrain dark matter models and gravitational wave emission \cite{cardos}.

In addition, black holes in an Anti-de Sitter (AdS) space-time can be thermally stable since the AdS boundary behaves as a reflecting wall to trap scattered waves and Hawking radiation \cite{ya}. Black holes as thermodynamical objects have an important role in gauge/gravity duality. According to  AdS/CFT correspondence, from the gravity side, a charged scalar field coupled to a charged black hole can make it unstable by the formation of a hair through superradiance scattering when Hawking temperature of the black hole drops below a certain critical temperature and spontaneously produces scalar hair with spherical symmetry. The emergence of a hairy black hole is related to the formation of a charged scalar condensation in the dual CFTs \cite{oscar}. In the dual field theory, this instability corresponds to a phase transition which in turn points to the spontaneous breakdown of the underlying gauge symmetry.

A Reissner-Nordstron (RN)-AdS black hole may also become unstable against a perturbing scalar field at low temperature when the effective mass squared becomes negative near the horizon. This is related to the near horizon scalar condensation instability, corresponding to what is known as a holographic superconductor in the context of AdS/CFT \cite{oscar,hart,bauso,bauso1,gary,va}. In such a context, the transition to superconductivity is described by a classical instability of planar black holes in an AdS space-time caused by a charged perturbing scalar field \cite{hart}. The holographic superconductivity was first investigated by Gubser \cite{gub} where it was concluded that  holographic superconductivity in a charged, complex scalar field around an AdS black hole is described by the mechanism of spontaneous $U(1)$ gauge symmetry breaking. It means that local symmetry breaking in the bulk corresponds to a global $U(1)$ symmetry breaking at the boundary on account of AdS/CFT correspondence. According to AdS/CFT dictionary, a condensate is described as a hairy black hole dressed with a charged scalar field in a holographic superconductor. It has numerically been found that the phase transition is of second order in a planar symmetric space-time and that there is a hairy black hole for $T<T_{c}$ \cite{pen,hor,pan}.

Born-Infeld (BI) electrodynamics as the non-linear extension of Maxwell electrodynamics was first presented in the 1930’s to develop a classical theory of charged particles with finite self-energy. However,  the emergence of quantum electrodynamics (QED) in later years and the accompanying renormalization program left the BI theory by the roadside \cite{la}. Nonetheless, the discovery of string theory and D-branes have revived it to some extent in recent years \cite{fra}.  It is recognized that BI electrodynamics appears in the low energy limit of string theory, encoding the low-energy dynamics of D-branes \cite{shi, re,de,wa}. The exact solutions of Einstein-Born-Infeld (EBI) gravity with zero \cite{gib} or nonzero cosmological constant \cite{wa,de}  and thermodynamic properties of these solutions have been studied in the past \cite{fer}.
In \cite{chen}, the effects of nonlinear electrodynamics on the holographic superconductors was investigated numerically by neglecting the back reaction of the scalar field on the metric. In \cite{shi}, the holographic superconductor in BI electrodynamics was studied by taking the back reaction of the scalar field on the background using the Sturm-Louville variational method which resulted in a relation between the critical temperature and charge density, showing that the critical temperature decreases by the growth of the BI coupling parameter, making the phase transition harder to occur. Their result was compatible with that obtained in \cite{chen}.

Our motivation is to investigate the effects of higher derivative gauge field terms (the nonlinear electrodynamics)
on superradiant instability, critical temperature and phase transition. In this paper, we first consider EBI-charged scalar field theory in an AdS space-time and review the equations of motion in section \ref{BHa}. We then  investigate instabilities of BI black holes under spherically symmetric charged scalar perturbations in section \ref{BHb} and move on to study static, spherically symmetric
black hole solutions with nontrivial charged scalar hair in section \ref{BH}. To see if these hairy black holes can be plausible endpoints of the charge superradiant instability, we study their stability under linear, spherically symmetric perturbations. Conclusions are drawn in the final sections.

\section{Setup and field equations \label{BHa}}
We study a system where gravity is minimally coupled to BI nonlinear electrodynamics and a massive charged scalar field in an AdS space-time. The action is given by
\begin{equation}\label{1}
  S=\int d^{4}x \sqrt{-g} \left[\frac{1}{2\kappa^2}(R-2\Lambda)+L_{BI}-g^{\mu \nu}D_{(\mu}^* \Phi^* D_{\nu)} \Phi-m_{s}^2 \Phi \Phi^*\right],
\end{equation}
where $\kappa^2=8\pi G$, $D_{\mu}=\nabla_{\mu}-iq A_{\mu}$, $A_{\mu}$ is the vector potential, $\Lambda=-\frac{3}{L^2}$ and $\Phi$ is a complex scalar field. The asterisk, $q$ and $m_{s}$ indicate complex conjugate, charge and mass of the scalar field with $\kappa^2=1$. The BI lagrangian is defined as
 \begin{equation}\label{2}
 L_{BI}=\frac{1}{b}\left(1-\sqrt{1+\frac{b F}{2}}\right),
\end{equation}
where $b$ is a BI coupling parameter, $F=F_{\mu \nu}F^{\mu \nu}$ and $F_{\mu \nu}=\nabla_{\mu}A_{\nu}-\nabla_{\nu}A_{\mu}$.

Varying action (\ref{1}) with respect to the metric, electromagnetic field and scalar field leads to the following equations of motion
\begin{eqnarray}\label{3}
&&R^{\mu \nu}-\frac{g^{\mu \nu} R}{2}-\frac{3 g^{\mu \nu}}{L^{2}}= \frac{g^{\mu \nu}}{b}\left(1-\sqrt{1+\frac{b F}{2}}\right)+\frac{F_{\sigma}^{\mu}F^{\nu \sigma}}{\sqrt{1+\frac{b F}{2}}}-g^{\mu \nu}m_{s}^{2} \Phi^{2}\nonumber\\
&&-g^{\mu \nu}| \nabla \Phi-i q A \Phi |^{2}+\left[(\nabla^{\nu}+i q A^{\nu})\Phi^{*}(\nabla^{\mu}-i q A^{\mu})\Phi+\mu \leftrightarrow \nu\right] ,
\end{eqnarray}
\begin{eqnarray}
&&(\nabla_{\mu}-i q A_{\mu})(\nabla^{\mu}-i q A^{\mu})\Phi-m_{s}^2\Phi=0,\label{4}\\
&&\nabla_{\mu}\left(\frac{F^{\mu \nu}}{\sqrt{1+\frac{bF}{2}}}\right)=i q \left[\Phi^{*}(\nabla^{\nu}-i q A^{\nu})\Phi-\Phi(\nabla^{\nu}+i q A^{\nu})\Phi^{*}\right],\label{5}
\end{eqnarray}
when $b\rightarrow 0$, the above equations reduce to the usual Einstein-Maxwell-scalar field theory.

At the linear level which implies small amplitude for the scalar field, it is reasonable to assume that the scalar field vanishes, so one may neglect the back-reaction of the scalar field on the electromagnetic and gravitational fields. For this reason, we use the following metric \cite{kru}
\begin{equation}\label{6}
ds^{2} =-V(r)dt^{2}+ \frac{dr^{2}}{V(r)}+r^{2}d\Omega^{2},
\end{equation}
where $V(r)$  takes the form \cite{kru}
\begin{equation}\label{7}
V(r)=1-\frac{M}{r}+\left[\frac{2}{3b}+\frac{1}{L^2}\right]r^2-\frac{2}{3b}\sqrt{r^4+b Q^2}+\frac{4 Q^2}{3 r^2} \times _{2}F_{1}\left[\frac{1}{4},\frac{1}{2},\frac{5}{4},-\frac{Q^2b}{r^4}\right].
\end{equation}
Here, $M$ and $Q$ are related to the mass and charge of the black hole and $_{2}F_{1}\left[\frac{1}{4},\frac{1}{2},\frac{5}{4},-\frac{Q^2b}{r^4}\right]$ is a hypergeometric function. By expanding convergent series of $_{2}F_{1}[a,b,c,z]$ for $|z|<1$ \footnote{$_{2}F_{1}[a,b,c,z]=\sum_{n=0}^\infty \frac{ (a)_n (b)_n}{(c)_n}\frac{z^n}{n!}$}, we find the behavior of $V(r)$ for large $r$  \cite{de}
\begin{equation}\label{7b}
V(r)=1-\frac{M}{r}+\frac{Q^2}{r^2}+\frac{r^2}{L^2}-\frac{ Q^4 b}{20 r^6}.
\end{equation}
It is seen that when $b\longrightarrow0$, $V(r)$ has the form of a Reisner-Nordstrom (RN)
AdS black hole. In \cite{kru}, a class of solutions of Eq. (\ref{5}) was presented as follows
\begin{equation}\label{8}
F_{r t}=\frac{Q}{\sqrt{r^4+b Q^2}}.
\end{equation}
Using Eq. (\ref{8}), we also compute the related gauge field
\begin{equation}\label{9}
A_{t}=\frac{Q}{r}\times_{2}F_{1}\left[\frac{1}{4},\frac{1}{2},\frac{5}{4},-\frac{Q^2b}{r^4}\right]-C,
\end{equation}
where $C$ is a constant of integration. Since we need to have $A_{t}(r_{+})=0$ \cite{yosh, 34, 35}, we take $C=\frac{Q}{r_{+}}\times_{2}F_{1}\left[\frac{1}{4},\frac{1}{2},\frac{5}{4},-\frac{Q^2b}{r_{+}^4}\right]$, where $r_{+}$ is the event horizon and $V(r_{+})=0$. As discussed in \cite{de, kru, tao}, when $V(r)=0$ there can be one or two horizons depending on the value of $M$.
\section{Instabilities of BI black holes\label{BHb}}
One may consider black hole instabilities in the context of AdS/CFT correspondence which is a powerful tool to study strongly coupled gauge theories using classical gravitation.  In an AdS space-time,  a static black hole corresponds to a thermal state in CFT on the boundary. So perturbing a black hole in an AdS space-time can be related to perturbing a thermal state in the corresponding CFT part.

When a BI black hole is perturbed by a scalar field, two types of linear instabilities with different physical nature can emerge. One is the superradiant instability in global small black holes, so-called when $r_+\ll L$, and the other is the near horizon scalar condensation instability which was first found in planar AdS black holes,  corresponding to a holographic superconductor in the context of AdS/CFT correspondence.
\subsection{ Small BI black holes and superradiant instability}
\begin{figure}[!ht]
\includegraphics[width=8.6cm,height=5cm]{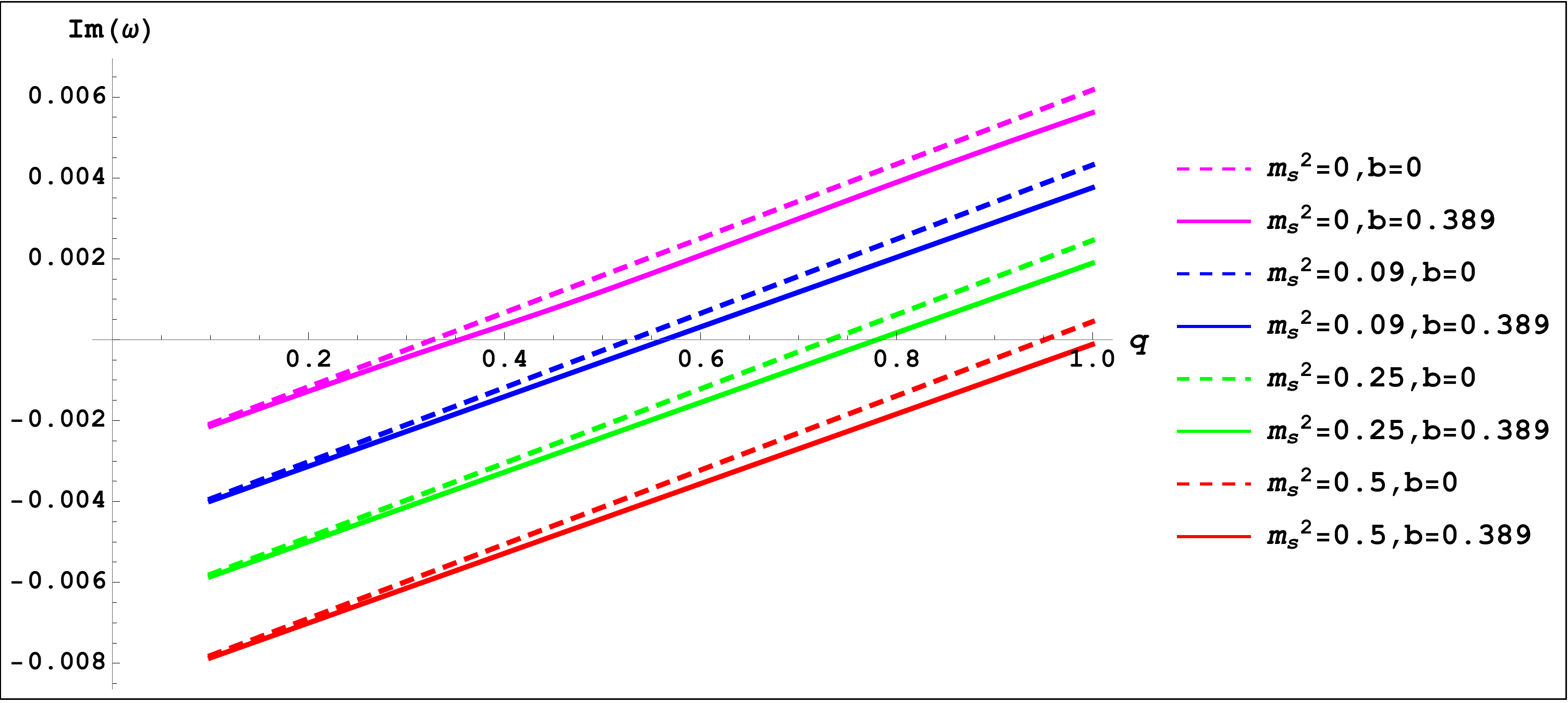}
\includegraphics[width=8.6cm,height=5cm]{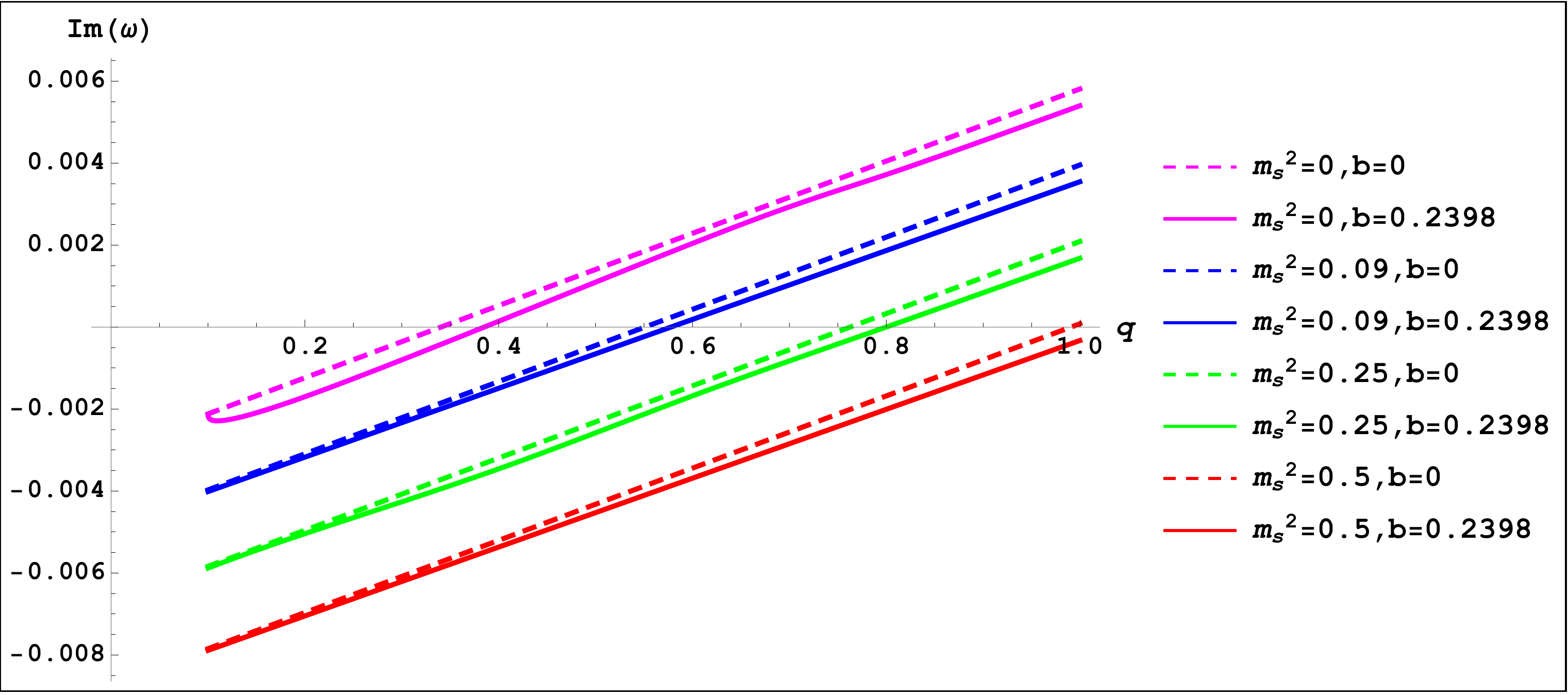}
\caption{The imaginary parts of the frequency as a function of the scalar charge for different values of scalar mass and $L=10$. Left: $Q=0.99$, $M=2$, the solid line is for $b=0.389$ and dashed line for $b=0$. Right: $Q=0.79$, $M=1.6$, the solid line is for $b=0.2398$ and dashed line for $b=0$.   }
\label{fig1}
\end{figure}

We consider a monochromatic and spherically-symmetric perturbation with frequency $\omega$
\begin{equation}\label{11}
 \Phi (r,t)= \frac{\psi(r)e^{-i\omega t}}{r}.
\end{equation}
Substituting the above ansatz to Eq. (\ref{4}), one gets
\begin{equation}\label{12}
 V^{2} \psi''+V V' \psi'+\left[\left(\omega + q A_{t}\right)^{2}-V\left(\frac{l(l+1)}{r^2}+m_{s}^{2}+\frac{V'}{r}\right)\right]\psi=0,
\end{equation}
where a prime indicates derivative with respect to $r$. By defining $\frac{dr_{*}}{dr}=\frac{1}{V}$, one obtains the asymptotic solutions of the scalar field
\begin{eqnarray}
&&\psi \sim e^{-i \hat{\omega} r_{*}}       \qquad \hat{\omega}=\omega + q A_{t}(r=r_{+}) \qquad r_{*}\rightarrow -\infty ,\label{13a}\\
 &&\psi\sim r^{-\frac{1}{2}\left(1+\sqrt{4m_{s}^{2}L^{2}+9}\right)}      \qquad    r_{*}\rightarrow +\infty.\label{13}
 \end{eqnarray}
We have analytically derived the real and imaginary parts of the frequency to lowest order in Appendix A, following the method presented in \cite{herd,card}, with the result
 \begin{eqnarray}\label{14}
&&\mbox{Re}(\omega)=\frac{3}{2L}+\sqrt{m_{s}^2+\frac{9}{4 L^2}}-q C  ,\nonumber\\
 &&\mbox{Im}(\omega) = -\frac{2r_{+}^2 \times\Gamma\left(\frac{3}{2}+\sqrt{m_{s}^2L^2+\frac{9}{4}}\right)}{  \Gamma(\frac{1}{2})\times\Gamma\left(1+\sqrt{m_{s}^2L^2+\frac{9}{4}}\right)} \times \frac{\mbox{Re}(\omega) }{L^2} .
\end{eqnarray}
To investigate superradiant instability, we consider solutions possessing two horizons, namely $r_{+}$ and $r_{-}$, which are obtained by setting specific values for $M$ \cite{de, kru, tao}. We define the superradiant regime as the onset of $\mbox{Re}(\hat{\omega})=\mbox{Re}({\omega})+q C-qQ/r_{+}<0$  which implies $\mbox{Im}(\omega)>0$ and the exponential growth of the scalar field wave function with time, leading to black hole instability. To analyze the instability numerically, we use the shooting method \cite{ra}, integrating Eq. (\ref{12})
numerically from $r_{+}$ to $L$, for which the base values are given by equations (\ref{14}). We repeat integrating by changing the base value of the frequency until $\psi(L)=0$. Fig. \ref{fig1} demonstrates $\mbox{Im}(\omega)$ as function of the scalar charge for different values of the scalar mass and two values of the BI coupling parameter (solid lines) which is compared to $b=0$ (dashed lines). As can be seen, $\mbox{Im}(\omega)$ is positive in some intervals, meaning that there is superradiant instability whose mode grows exponentially. On the other hand, this figure shows that superradiant instability has an inverse relation with the scalar mass and BI coupling parameter. The larger the BI coupling parameter the slower the growth rate of the  instability. As can be seen in Fig. \ref{fig1}, $\mbox{Im}(\omega)$ for small $q$ at linear and nonlinear electrodynamics has a similar value approximately, since the coupling between the charged scalar field and the gauge field becomes weaker.
\subsection{ Large BI black holes and near horizon scalar condensation instability}
Let us consider action (\ref{1}) and Klein-Gordon equation (\ref{12}). In a charged black hole background (BI black hole), the charged scalar field gains an effective mass square
\begin{eqnarray}\label{15a}
m_{eff}^{2}=m_{s}^{2}- \frac{q^{2} A_{t}^{2}}{V}.
\end{eqnarray}
Near the event horizon, $V$ is very small, so $m_{eff}^{2}$ can be negative enough to destabilize the scalar field. Assuming back reaction of the scalar field on geometry,  the scalar field becomes unstable on the background which would lead to a hairy black hole solution. Such unstable modes live in an asymptotically-AdS space-time and are very important to the formation of hairy black holes. When the black hole is near extremality, such instabilities become more pronounced, since in addition to $V$, $V'$ also becomes approximately zero, so that $\frac{1}{V}$ diverges faster. Such an unstable mode is associated with the near horizon geometry of an extremal charged black hole and if it satisfies  $m_{eff}^{2}<m_{BF}^{2}=-\frac{9}{4 L^2}$ \cite{va}, where $m_{BF}$ is the Breitenlohner-Freedman bound of the near horizon, it becomes space-like, creating a tachyonic instability. Also, an instability of this kind exits in the non-extremal black holes provided that $q$ becomes large enough.
\section{Hairy black hole solutions in Einstein- Born- Infeld- scalar field theory \label{BH}}
Is it conceivable that the possible end states of instabilities discussed above become a hairy black hole that has a charged scalar condensate near the horizon? We associated a small hairy black hole, $r_+\ll L$, with the endpoint of superradiant instability and a planar hairy black hole with the near horizon scalar condensation instability. In what follows we seek to answer the above question.
\subsection{Equation of motion at nonlinear level}\label{BH1}
To obtain hairy solution near critical temperature, we consider the ansatz
\begin{eqnarray}\label{15}
ds^{2} =-V(r)h(r)dt^{2}+ \frac{dr^{2}}{V(r)}+r^{2}d\Omega^{2}_{k},
\end{eqnarray}
where $V(r)=k-\frac{2m(r)}{r}+\frac{r^2}{L^2}$ with $k=0,1$ where $k=0$ is related to black holes with planar horizon and $k=1$ to black holes with spherical horizon. The function $h(r)$ shows matter field backreaction effects on the space-time geometry \cite{win, ra}. Now, use of the gauge freedom \cite{win} renders the scalar field real and dependent on radial coordinate $\Phi=\psi(r)$ and we take the vector potential $A(r)=\phi(r) dt$. Consequently, using equations (\ref{3})-(\ref{5}) leads us to the following four nontrivial and coupled equations
\begin{equation}
V'-\left(\frac{3r}{L^2}-\frac{V}{r}+\frac{k}{r}\right)+r\left[m_{s}^{2}{\psi}^{2}-\frac{1}{b}+\frac{1}{b\sqrt{1-\frac{b{\phi'}^2}{h}}}+V\left({\psi'}^2+\frac{q^2 \phi^2 \psi^2}{V^2 h}\right)\right]=0,\label{17}
\end{equation}
\begin{eqnarray}
&&\frac{h'}{h}=2 r\left({\psi'}^{2}+\frac{q^2 \phi^2 \psi^2}{V^2 h}\right),\label{16}\\
&&\phi''+\frac{2\phi'}{r}\left(1-\frac{b{\phi'}^{2}}{h}\right)-\frac{\phi' h'}{2 h}-\frac{2q^2\psi^2\phi}{V}{\left(1-\frac{b\phi'^2}{h}\right)}^{\frac{3}{2}}=0,\label{18}\\
&&\psi''+\left(\frac{2}{r}+\frac{V'}{V}+\frac{h'}{2h}\right)\psi'+\left(\frac{q^2 \phi^2}{V^{2}h}-\frac{m_{s}^2}{V}\right)\psi=0.\label{19}
\end{eqnarray}
The above nonlinear equations are not amenable to analytical solutions. To solve them numerically, we use the shooting method and integrate the coupled equations (\ref{16})-(\ref{19})
from $r_{+}$ to the reflective boundary. The regularity conditions for equation (\ref{18}) signify $\phi_{+}=0$ which satisfies the vanishing gauge field at the event horizon.  The boundary conditions at the event horizon, using equations (\ref{16})-(\ref{19}) become
\begin{eqnarray}
&&V'_{+}=\frac{3 r_{+}}{ L^2}+\frac{k}{r_{+}}-r_{+}\left[m_{s}^{2}\psi_{+}^{2}-\frac{1}{b}+\frac{1}{b\sqrt{1-\frac{b{\phi'_{+}}^2}{h_{+}}}}\right],\label{20}\\
&&\psi'_{+}=\frac{m_{s}^{2}\psi_{+}}{V'_{+}},\label{21}\\
&&h'_{+}=2h_{+}r_{+}\left[{\psi'_{+}}^2+\frac{q^{2}{\phi'_{+}}^{2}\psi_{+}^{2}}{{V'_{+}}^{2}h_{+}}\right].\label{22}
\end{eqnarray}
In addition, there is a reflective boundary condition that induces vanishing of the scalar field at the AdS boundary \cite{bosch}.

At this point, it would be interesting to investigate the relation between the above solutions and the concept of Hawking temperature. Hawking temperature of a black hole is given by $T_{H}=\frac{V'_{+}\sqrt{h_{+}}}{4 \pi}$. The choice $h(r\longrightarrow\infty)=1$, makes Hawking temperature the same as temperature of the field theory at the boundary \cite{hart}.
Now, according to AdS/CFT dictionary,  a mapping exists between CFT operators at the boundary and gravitational field in the bulk, that is, an operator $\mathcal{O}$ could be dual to a charged scalar field, say $\psi$, in the bulk. Therefore, to investigate phase transition, we consider the asymptotic behavior of the scalar field which is determined by $\psi=\frac{\psi_{1}}{r^{\Delta_{-}}}+\frac{\psi_{2}}{r^{\Delta_{+}}}$ with $\Delta_{\mp}=\frac{3}{2}\mp\sqrt{\frac{9}{4}+m_{s}^{2}L^2}$ being the dimension of the scalar operator \cite{hart}, where $\psi_{1}$ is considered as a source for the operator dual to $\psi$ and $\psi_{2}$ is its expectation value. For $m_s^2 L^2 \ge -\frac{5}{4}$, only $\psi_2$ mode is normalizable and the boundary condition to consider is $\psi_1=0$\footnote{For $-\frac{9}{4} \le m_s^2 L^2<-\frac{5}{4}$, $\psi_{1}$ and $\psi_2$ are both normalizable. To have a stable theory, we can have a choice of boundary conditions and may impose either $\psi_1=0$ or $\psi_2=0$ \cite{hor}.}  \cite{hor}. When $\psi_2 \propto \langle\mathcal{O}\rangle\ne 0$, the charged scalar operator $\mathcal{O}$ condenses and leads to the breaking of $U(1)$ global symmetry \cite{sym}. To break symmetry spontaneously, the scalar operator can condense without being coupled to an external source \cite{hor}. So we seek solutions that represent states of the conformal field theory without source and  set $\psi_{1}=0$. The phase transition can then be seen  by drawing plots of $\Delta h-T$, $\Delta h=h(\infty)-h(r_+)$, and $\psi_{2}-T$ \cite{oscar,hart}.
\subsection{Hairy black holes and numerical results}
\begin{figure}[!ht]
\includegraphics[width=8cm,height=5cm]{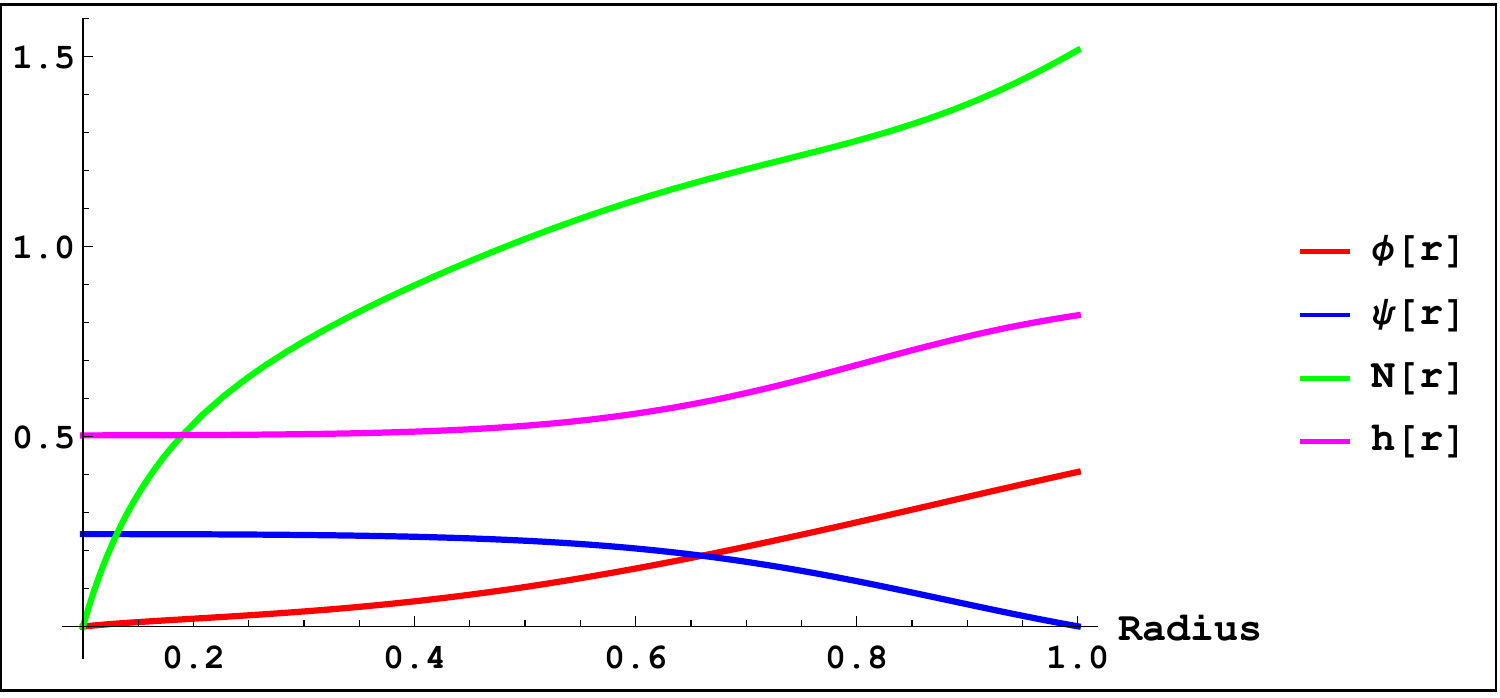}
\includegraphics[width=8cm,height=5cm]{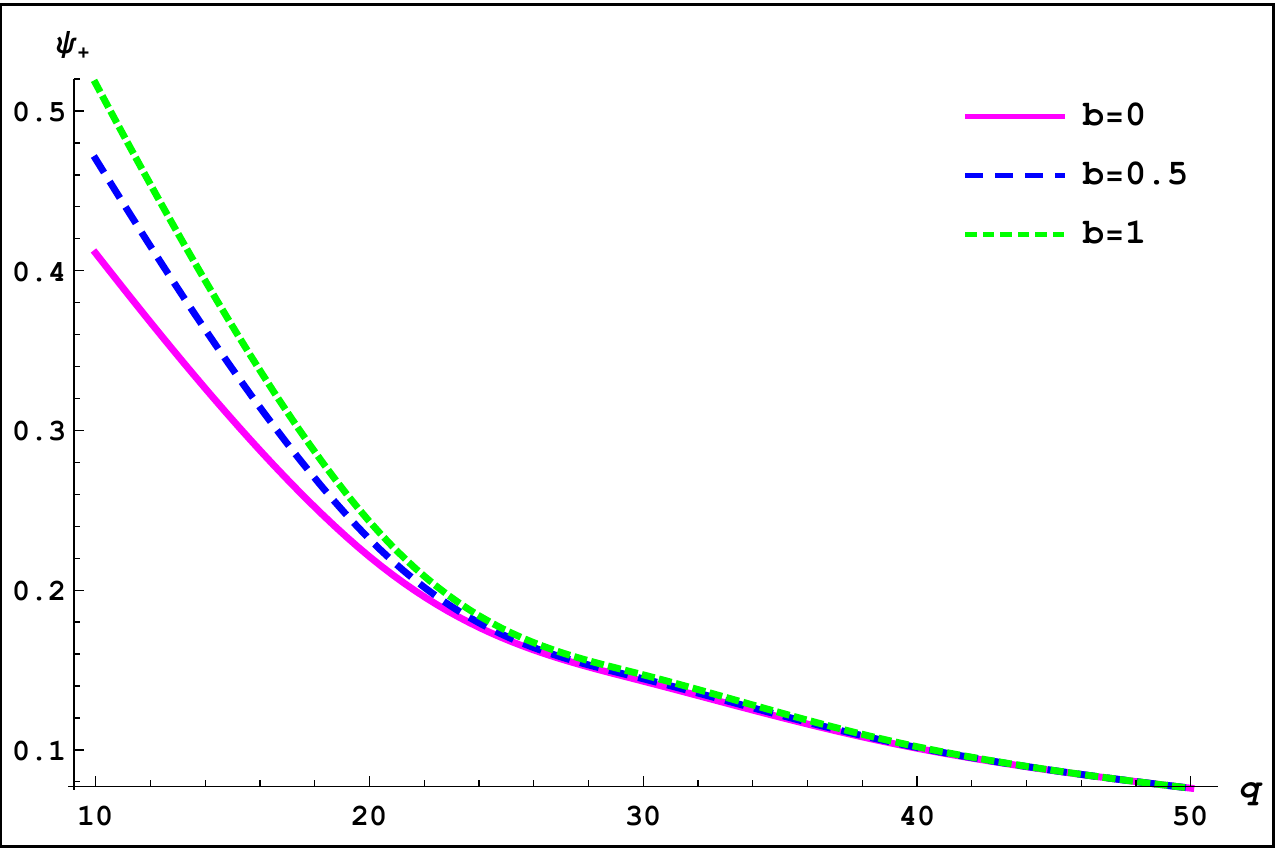}
\caption{Left: Field variables as a function of radius with $r_{+}=0.1$, $b=1$, $q=20$, $m_{s}^{2}=0$, $\phi'_{+}=0.3$ and $\psi_{+}=0.2432$. Right: $\psi_{+}$ is plotted as a function of $q$ when the scalar field has only one node at the reflective boundary with $\phi'_{+}=0.3$, $r_{+}=0.1$, $m_{s}^{2}=0$ and different values of $b$.  }
\label{fig2}
\end{figure}

\begin{figure}[!ht]
\includegraphics[width=8cm,height=5cm]{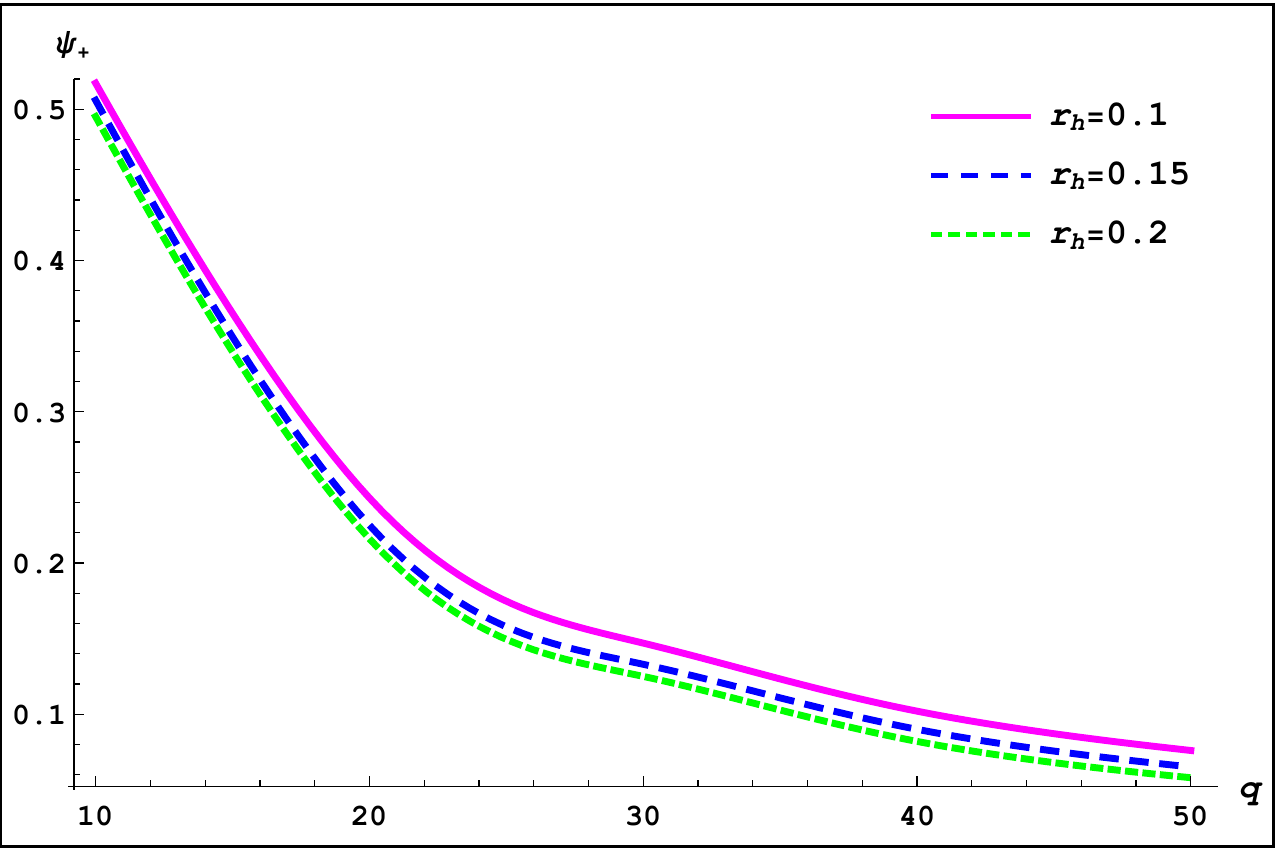}
\includegraphics[width=8cm,height=5cm]{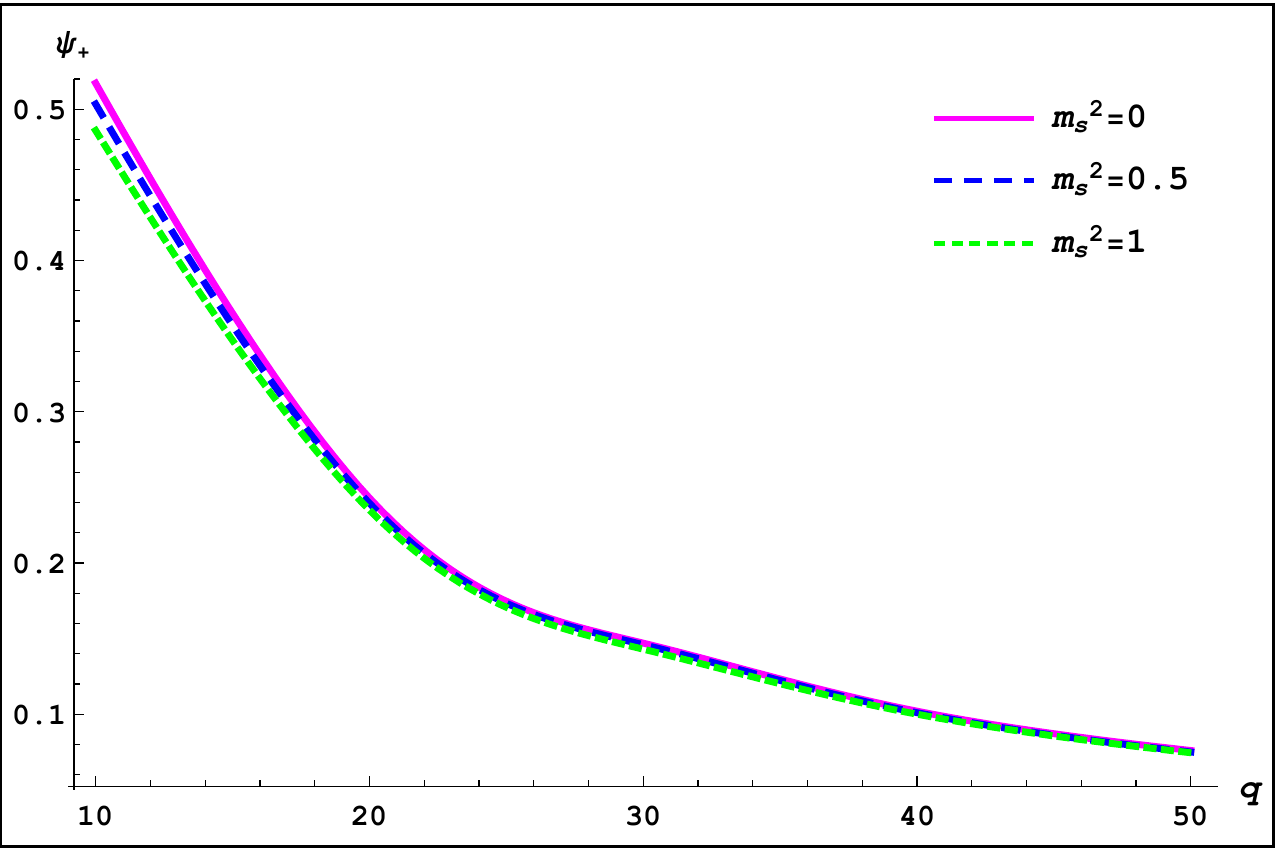}
\caption{$\psi_{+}$ is plotted as a function of $q$ when the scalar field has only one node at the reflective boundary with $\phi'_{+}=0.3$, $b=1$, Left: $m_{s}^{2}=0$ and different values of $r_{+}$. Right: $r_{+}=0.1$ and different values of $m_{s}^{2}$. }
\label{fig3}
\end{figure}

\begin{figure}[!ht]
\includegraphics[width=8cm,height=5cm]{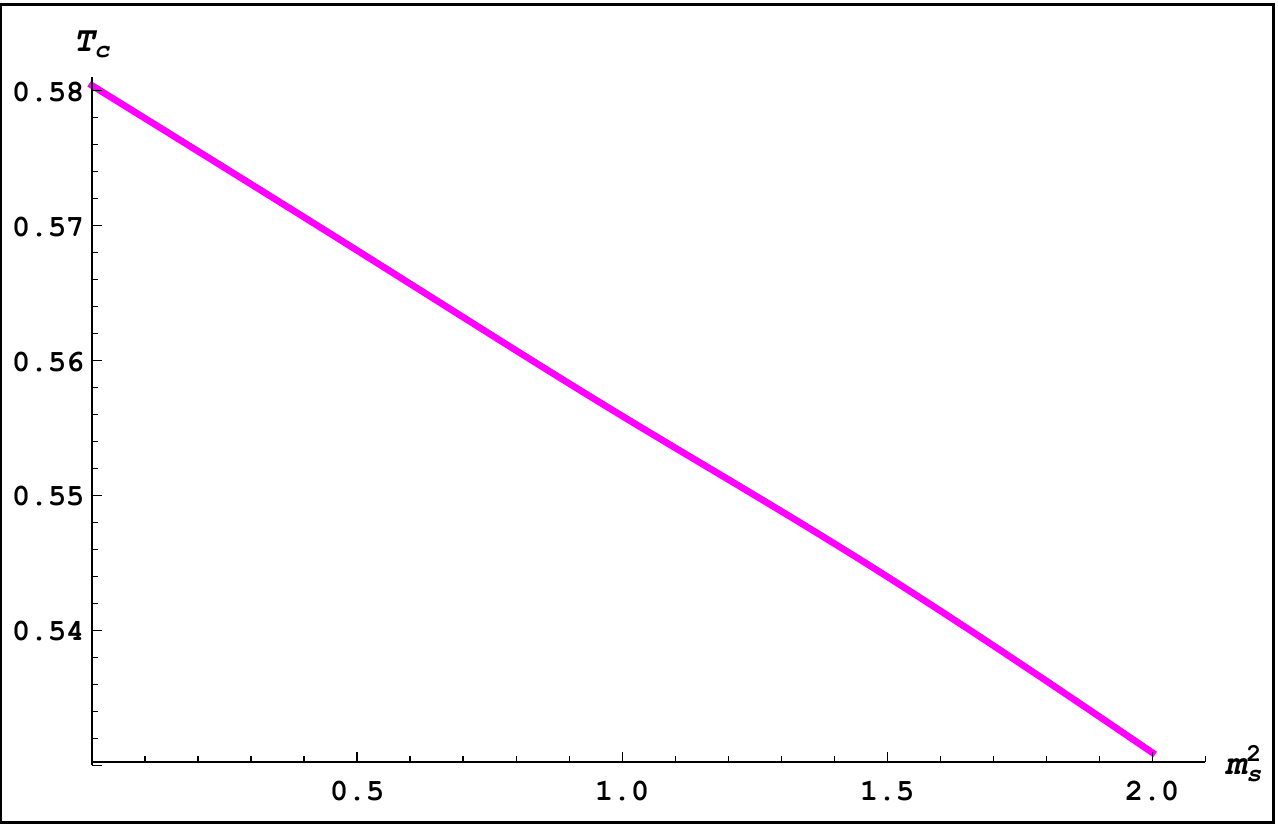}
\includegraphics[width=8cm,height=5cm]{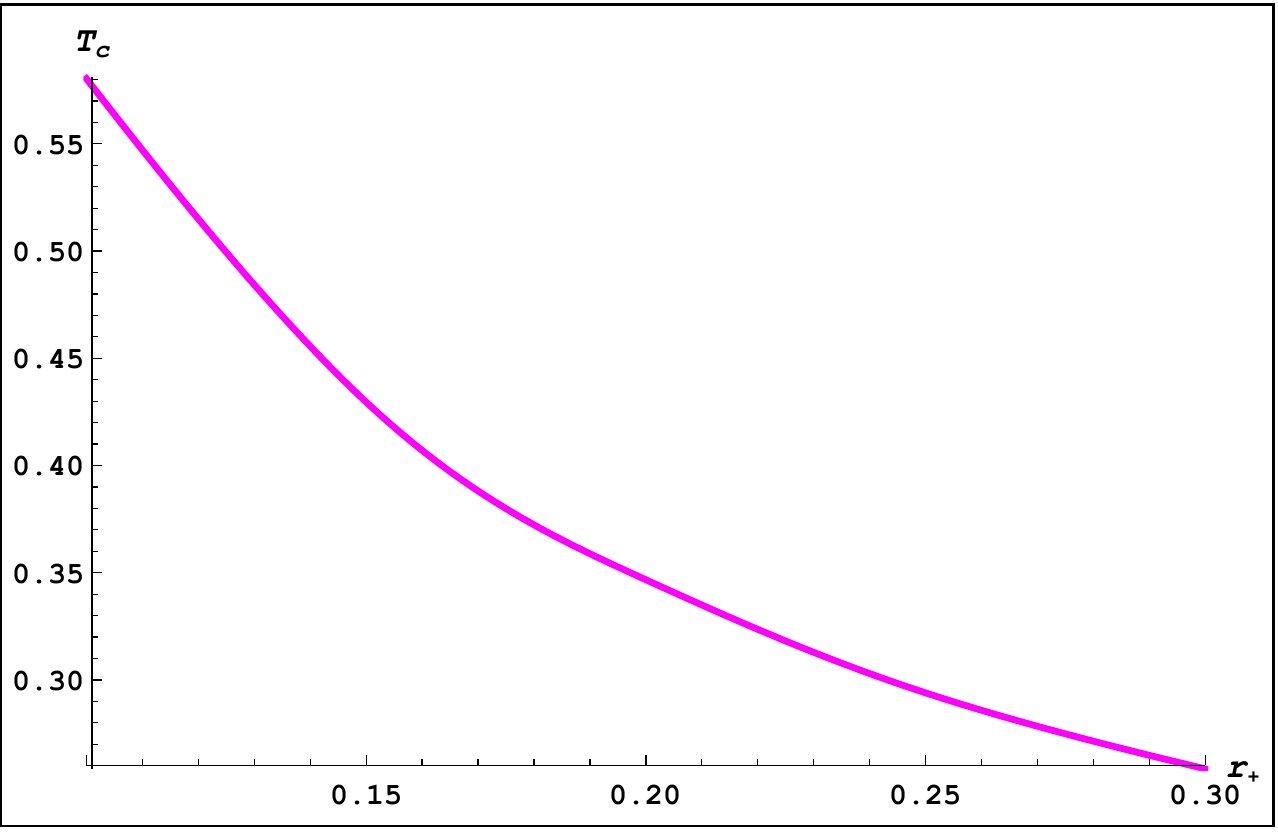}
\caption{The critical temperature is plotted as a function of, Left: $m_{s}^{2}$ and $r_{+}=0.1$. Right: the event horizon and $m_{s}^{2}=0$ with $q=20$, $b=1$.  }
\label{fig4}
\end{figure}

\begin{figure}[!ht]
\includegraphics[width=8cm,height=5cm]{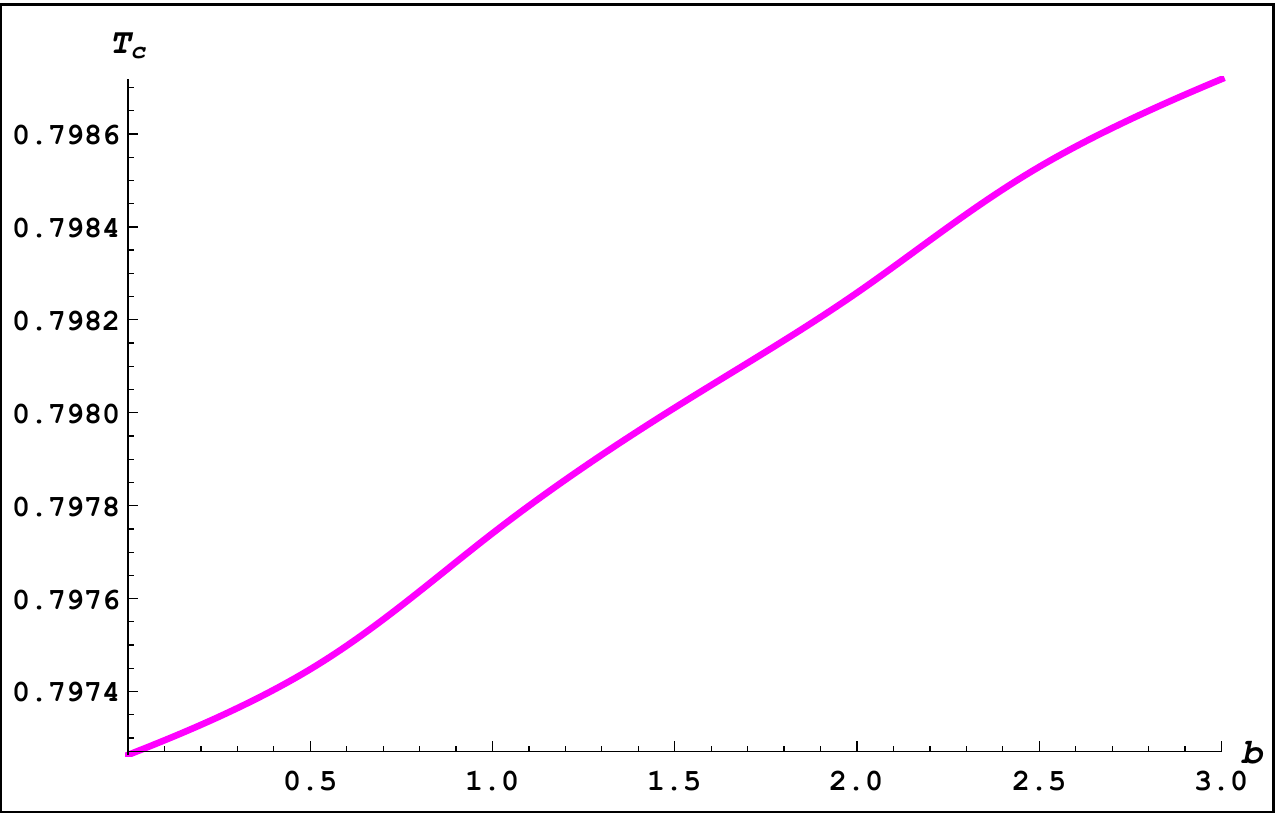}
\includegraphics[width=8cm,height=5cm]{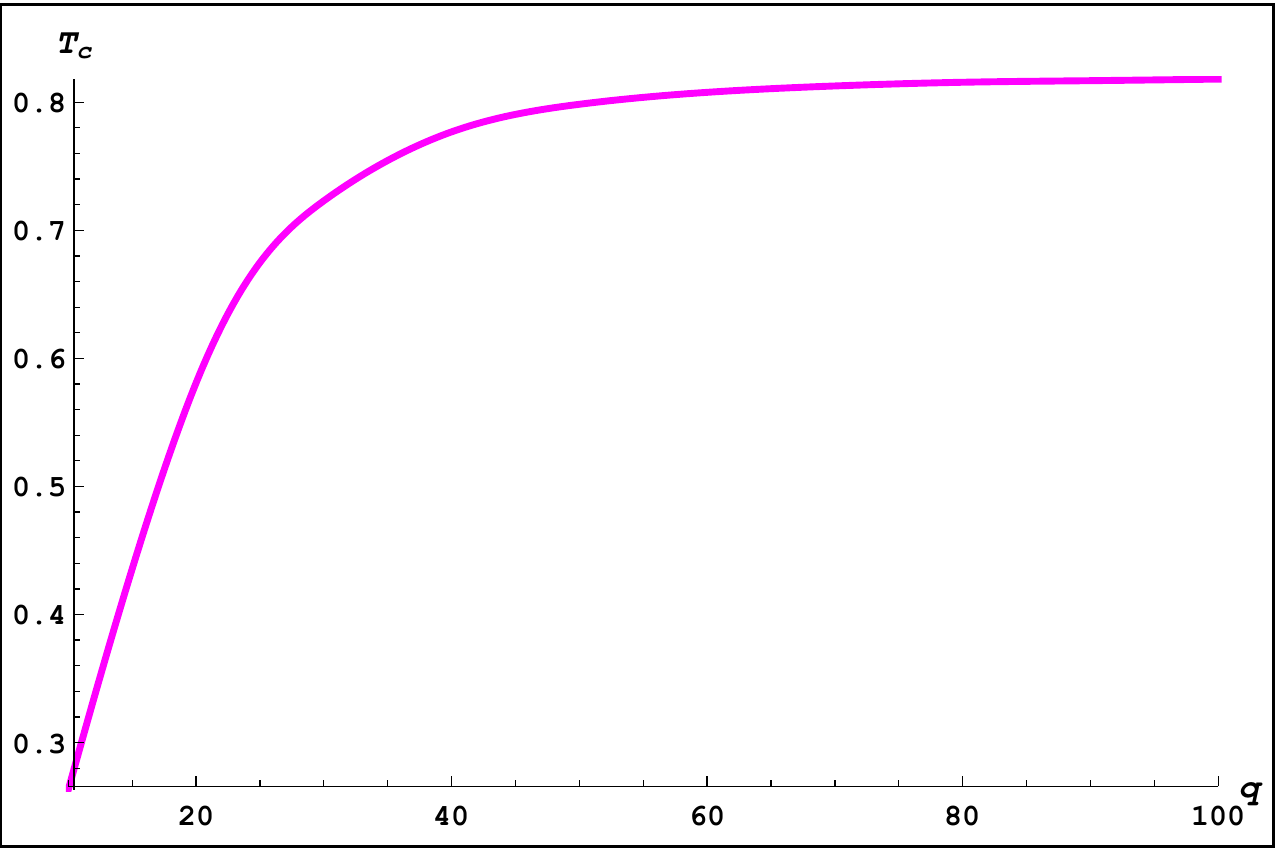}
\caption{The critical temperature as a function of, Left: the BI coupling parameter and $q=50$. Right: $q$ and $b=1$ with $r_{+}=0.1$, $m_{s}^{2}=0$, $\phi'_{+}=0.3$ and $\psi_{+}=0.2432$.  }
\label{fig4b}
\end{figure}

\begin{figure}[!ht]
\includegraphics[width=8cm,height=5cm]{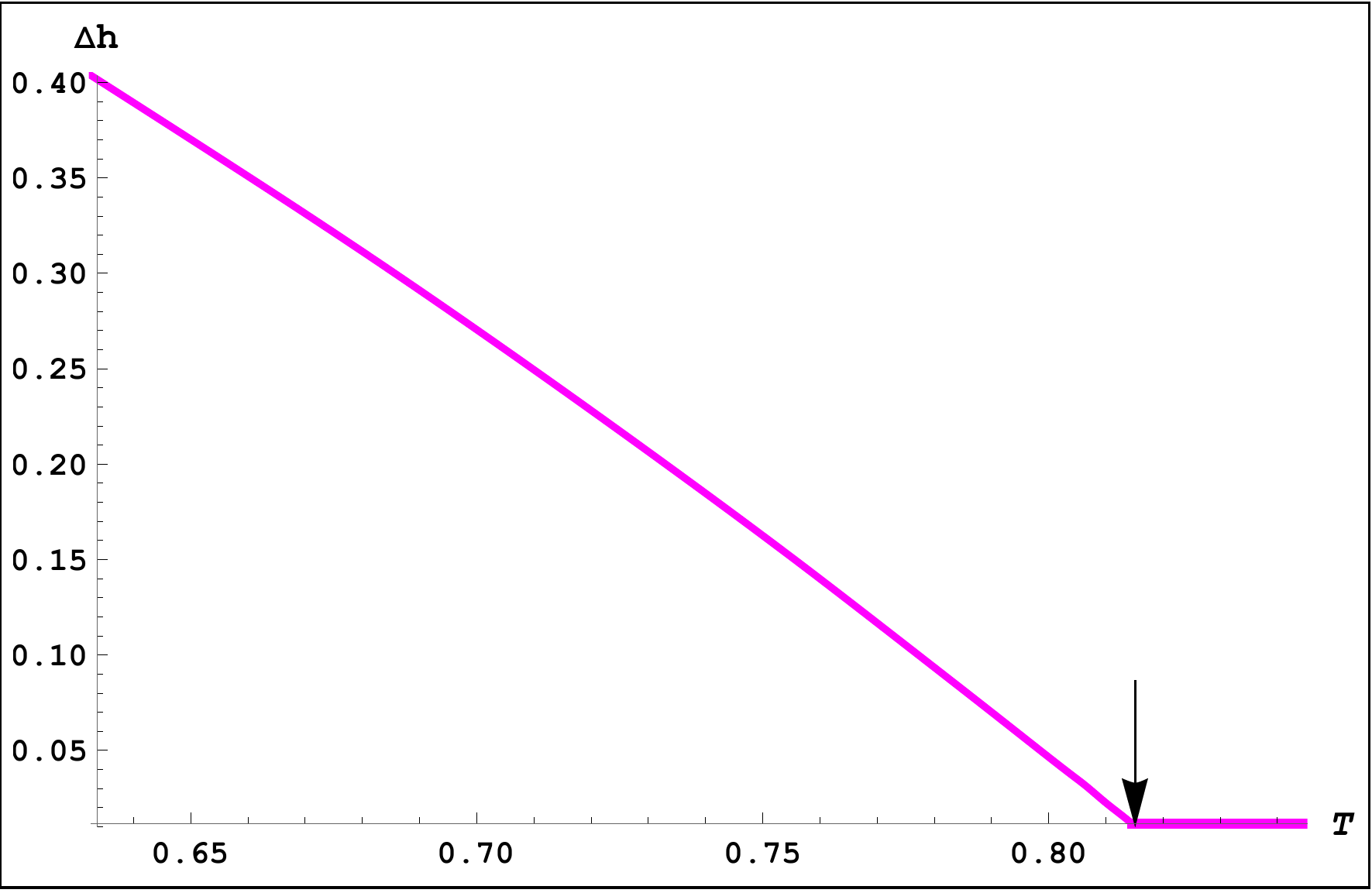}
\includegraphics[width=8cm,height=5cm]{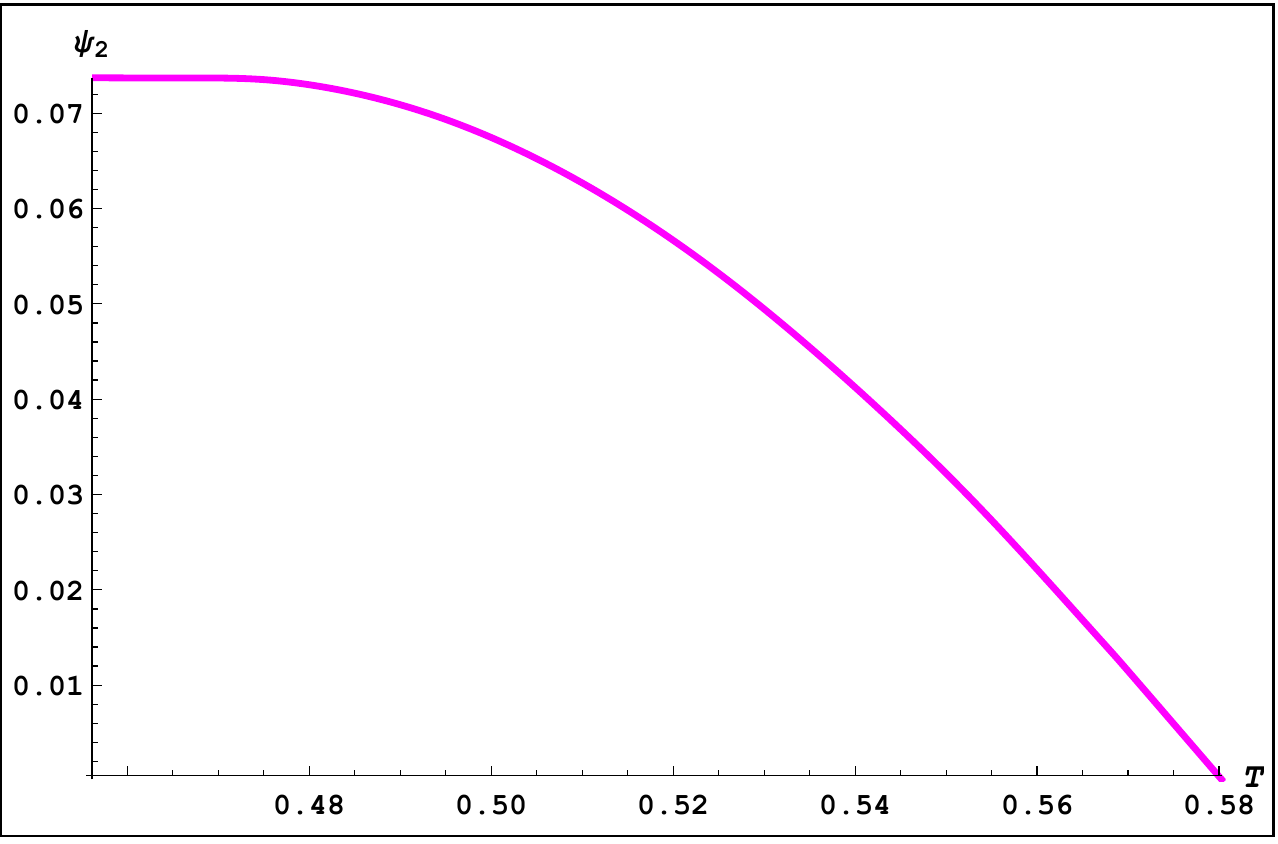}
\caption{ Left: $\Delta h$ as a function of temperature for $q=80$, the arrow signifies the critical temperature. Right: Parameter $\psi_{2}$ is plotted as a function of temperature for $q=20$ with $r_{+}=0.1$, $m_{s}^{2}=0$ and $b=1$.  }
\label{fig5}
\end{figure}

\begin{figure}[!ht]
\includegraphics[width=8cm,height=5cm]{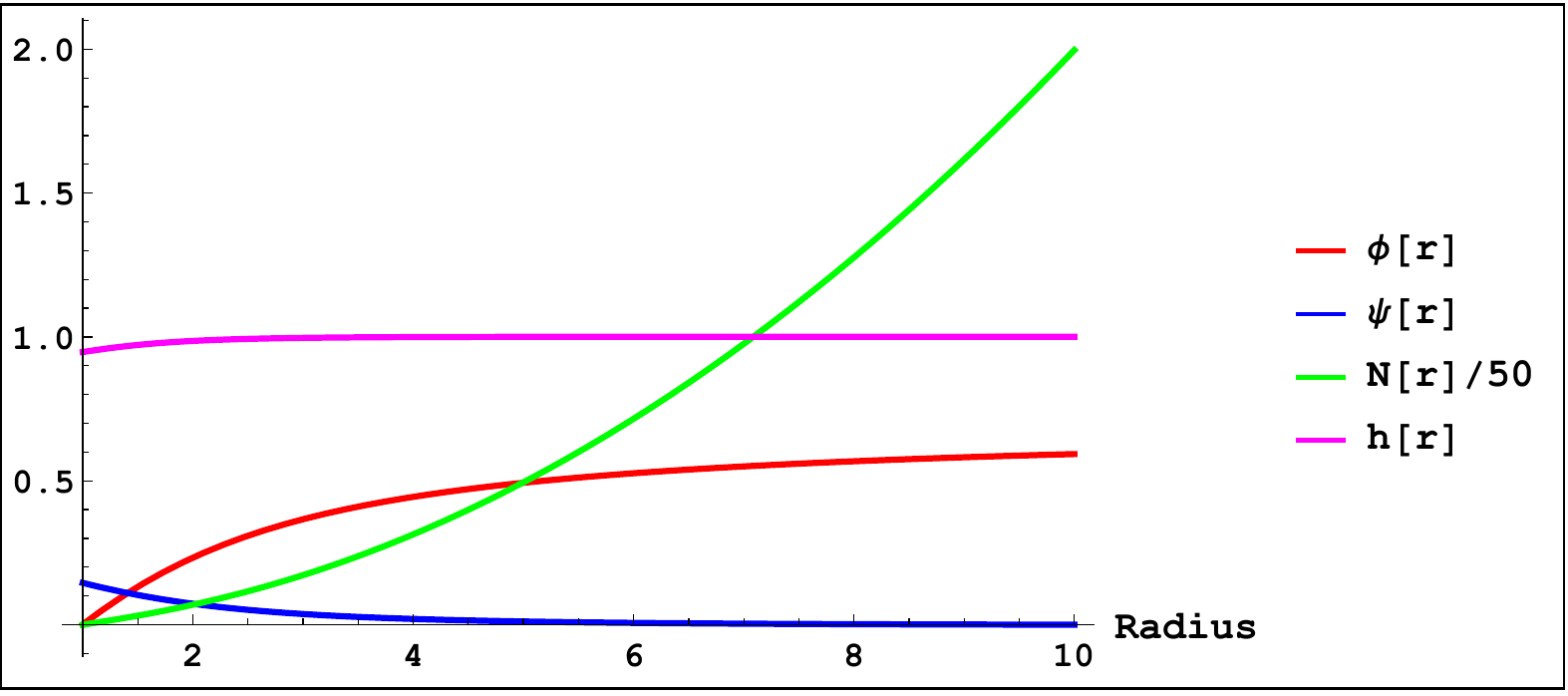}
\includegraphics[width=8cm,height=5cm]{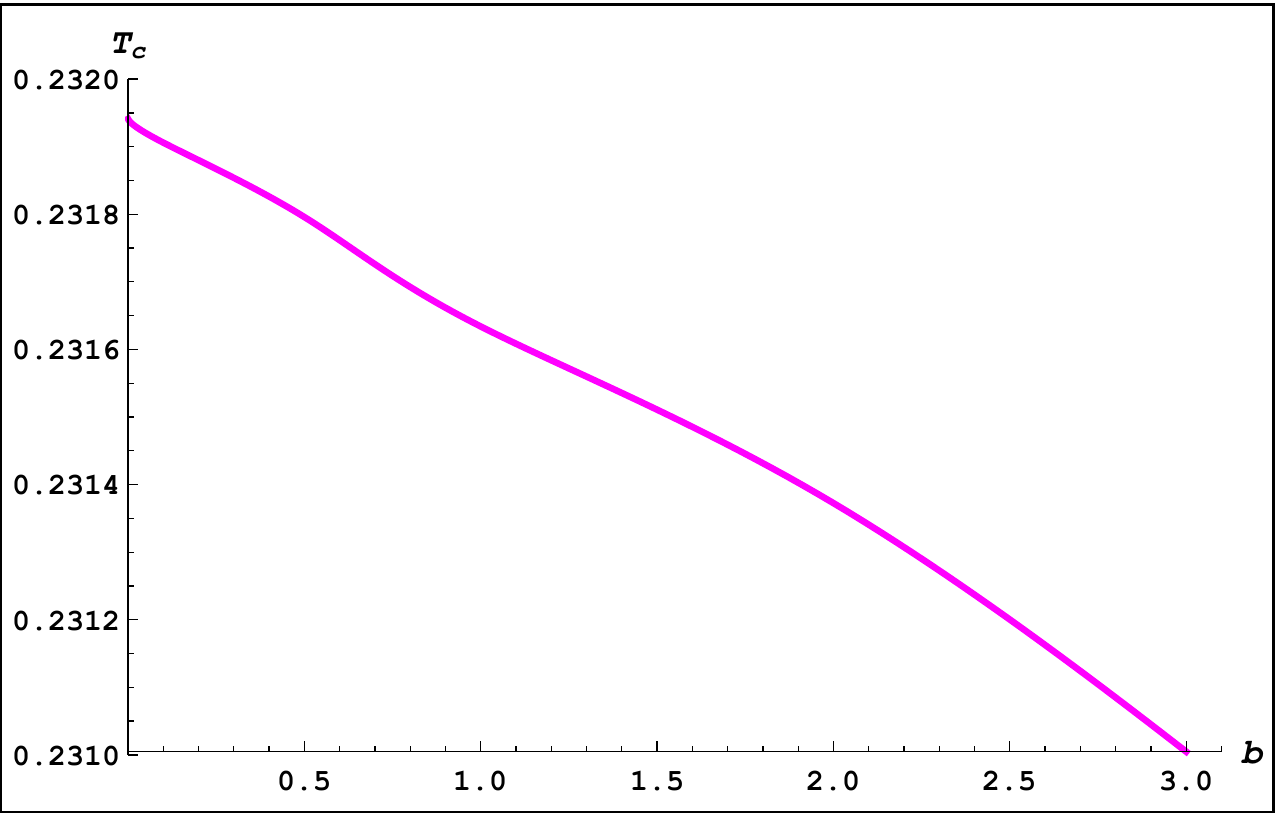}
\caption{Left: Field variables as a function of radius with $b=1$, $q=10$ and $\psi_{+}=0.1455$. Right: The critical temperature as a function of BI coupling parameter with $q=10$,  $\phi'_{+}=0.3$ and $m_{s}^{2}=-2$ for a planar hairy black hole.  }
\label{fig6}
\end{figure}

To study a small hairy black hole it is useful to use scaling symmetry as follows \cite{oscar,yao}
\begin{equation}\label{22a}
(t,r,\theta,\varphi)\rightarrow (\lambda t,\lambda r,\theta,\varphi),\qquad(V,h,\phi,\psi)\rightarrow (V,h,\phi,\psi),\qquad (q,L,r_{+})\rightarrow (\frac{q}{\lambda},\lambda L,\lambda r_{+}).
\end{equation}
Equations of motion then become invariant under the above scaling symmetry. We fix $L=1$ and use it to make quantities dimensionless. The left panel in Fig. \ref{fig2} shows that there is regular and nonsingular solutions outside the horizon in Einstein-Born-Infeld-scalar field theory. The right panel in Fig. \ref{fig2} and Fig. \ref{fig3} demonstrate the solution space for different values of the free parameters, namely $q$, $b$, $r_{+}$ and $m_{s}^{2}$. As can be seen in the right panel in Fig. \ref{fig2}, $\psi_{+}$ has a direct relation to $b$ for small scalar charges. However, for larger scalar charges, $\psi_{+}$ is essentially unchanged for different values of $b$. The left panel in Fig. \ref{fig3} shows that $\psi_{+}$ has an inverse relation to $q$ and $r_{+}$ which means that the smaller the value of  $q$ and $r_{+}$, the larger the value of $\psi_{+}$. In the right panel in Fig. \ref{fig3}, $\psi_{+}$ is plotted as a function of $q$ for different values of the scalar mass. As can be seen, for small scalar charges, $\psi_{+}$ has an inverse relation to the scalar mass, but for large scalar charges, the scalar mass has no significant effect on $\psi_{+}$. In Figs. \ref{fig4} and \ref{fig4b}, the dependence of $T_{c}$ on $m_{s}^{2}$, $r_{+}$, $b$ and $q$ is considered. $T_{c}$ being the critical temperature and the point at which $\psi$ vanishes. In addition, one may show that by an appropriate choice of the parameters, phase transition of a BI black hole to a hairy one occurs at $T_{c}$,  and that a hairy solution together with a solution akin to a BI-AdS black hole for which $\psi$ vanishes also exists. Note that one may consider $T_{c}$  as a starting point for the formation of the hairy black hole and then by continuously tuning the parameters there is a critical point in the parameter space beyond which $\psi$ is forced to have zero expectation value.  As can be seen, $T_{c}$ has an inverse relation to $m_{s}^{2}$ and $r_{+}$ and a direct relation to $b$ and scales with $q$ as $T_{c}\propto \sqrt{q}$. The left Fig. \ref{fig4b} implies that larger BI couplings lead to larger critical temperature and a more probable phase transition. Fig. \ref{fig5} shows variation of the critical temperature with solutions of the metric, $\Delta h$, and condensation $\psi_2$. As can be seen in the left panel, there is a hairy black hole for $T<T_{c}$ and a BI black hole for $T>T_{c}$ so that phase transition between the hairy black hole and BI black hole occurs at $T_c$. The right panel shows the existence of the condensation for $T<T_{c}$.\\
For a planar hairy black hole, in addition to $L$, we fix $r_{+}=1$ without loss of generality\footnote{A additional scaling symmetry exits for asymptotically local solutions, i.e. planar solutions, that allows us to have $r_+=1$ and $L=1$ simultaneously. }. In Fig. \ref{fig6}, we show the field variables of a planar hairy black hole as a function of radius, left panel, and the critical temperature as a function of the BI coupling parameter, right panel, for planar hairy black holes. It can be seen that critical temperature decreases as the BI coupling parameter increases, having almost a linear relationship to the BI coupling and making the onset of phase transition less probable. This result for the planar hairy black hole is different from that for a small hairy black hole which can be related to black hole geometry.
\section{ End point of superradiant instability and stability analysis\label{BHc}}
In this section we study the endpoint of superradiance instability which could be considered as a small hairy black hole, being stable at $T\sim T_{c}$\footnote{We show that the small hairy black hole is stable near $T_c$ when the scalar field has only one node at the AdS boundary, the ground state hairy black hole, while for $T<T_c$ the scalar field has more than one node which causes the hairy black holes to become unstable and is likely to decay into the ground state hairy black hole \cite{ra}.\\
The question may arise as to if one starts with a BI black hole at $T\ll T_c$, would such a BI black hole end up as a hairy BH at $T <T_c$ or $T \sim T_c$?  The answer lies in thermodynamic quantities of BI-AdS and hairy black holes which need to be defined and evaluated to gain a deeper understanding of phase transition properties and thermodynamic stability of such black holes. Therefore the free energy of a hairy black hole and a BI black hole should be evaluated numerically and compared at the same interval ($T<T_c$). The dominant phase is the one which has less free energy  \cite{bauso1}.
}. To investigate the system stability, we first consider its time evolution by assuming time dependence of the field variables in addition to radial
dependency and derive perturbation equations by linearly perturbing the system around static solutions \cite{ra}.
\subsection{Dynamical equations }
Defining $\xi=V\sqrt{h}$, dynamical equations become
\begin{equation}\label{22}
\frac{V'}{ V}+\frac{V-1}{Vr}-\frac{3r}{VL^2}=-\frac{r}{\xi^2}\left[|\dot{\psi}|^2+|\xi \psi'|^2+q^2|\phi|^2|\psi|^2+2q\phi\mbox{Im}(\psi\dot{\psi}^*)
-\frac{Vh}{b}\left(1-\frac{1}{\sqrt{1-\frac{b \phi'^2}{h}}}\right)+m_{s}^2V h \psi^{2}\right],
\end{equation}
\begin{eqnarray}
&&\frac{h'}{h}=\frac{2r}{\xi^2}\left(|\dot{\psi}|^2+|\xi \psi'|^2+q^2 |\phi|^2|\psi|^2+2q\phi\mbox{Im}(\psi\dot{\psi}^*)\right),\label{23}\\
&&
 \frac{ \xi'}{\xi}-\frac{3r}{VL^2} =\frac{r}{Vb}\left(1-\frac{1}{\sqrt{1-\frac{b \phi'^2}{h}}}\right)-\frac{r m_{s}^2 \psi^{2}}{V}+\frac{(1-V)}{Vr} ,\label{24}\\
&&
-\frac{\dot{V}}{V}=2r \mbox{Re}(\dot{\psi}^*\psi')+r q \phi\mbox{Im}(\psi'^*\psi),\label{25}
\end{eqnarray}
where a dot signifies derivative with respect to time. From Maxwell equations (\ref{5}), we get two dynamical equations
\begin{eqnarray}
&&\phi''+\frac{2\phi'}{r}\left(1-\frac{b\phi'^{2}}{h}\right)-\frac{\phi'h'}{2h}+\left(\frac{2q\mbox{Im}(\dot{\psi}\psi^*)}{V}-\frac{2q^2|\psi|^2 \phi}{V}\right)\left(1-\frac{b\phi'^{2}}{h}\right)^{\frac{3}{2}},\label{26}\\
&&
\partial_{t}{\left(\frac{\phi'}{h^{\frac{1}{2}}\left(1-\frac{b\phi'^{2}}{h}\right)^{\frac{1}{2}}}\right)}=-2q\mbox{Im}(\xi \psi'\psi^*).\label{27}
\end{eqnarray}
By defining $\psi=\frac{\Psi}{r}$, the Klein-Gordon equation (\ref{4})
is given by
\begin{equation}\label{28}
-\ddot{\Psi}+\left(\frac{\dot{\xi}}{\xi}+2iq\phi\right)\dot{\Psi}+\xi(\xi \Psi')'+\left(iq\dot{\phi}-\frac{\xi \xi'}{r}-iq\frac{\dot{\xi}}{\xi}\phi+q^2\phi^2-\frac{\xi^2}{V} m_{s}^2\right)\Psi=0.
\end{equation}
\subsection{Perturbation equations }
Let us now linearly perturb the system around static solutions as $V(r,t)=\bar{V}+\delta V(r,t)$,  where $\bar{V}$ and $\delta V(r,t)$ show static solutions and perturbations, and similarly for other field variables and substitute them in equations (\ref{22}-\ref{28}). Now, defining $\delta \Psi =\delta u+i \delta \dot{w}$  and eliminating metric variables, perturbation equations are obtained as two dynamical equations and a constraint \cite{ra}
\begin{eqnarray}\label{29}
&&\delta \ddot{u}-\bar{\xi}^2 \delta u''-\bar{\xi} \bar{\xi}'\delta u'+\left[3 q^2 \bar{\phi}^2+\frac{\bar{\xi} \bar{\xi}'}{r}+2 \bar{V}\left(\frac{\bar{\Psi}}{r}\right)'^2 \left(\frac{r^2 \bar{h}}{b{\left(1-\frac{b\bar{\phi}'^2}{\bar{h}}\right)}^{\frac{1}{2}}}-\frac{r^2 \bar{h}}{b}-\frac{3r^2 \bar{h}}{L^{2}}-\bar{h}\right)+m_{s}^{2}\bar{V}\bar{h}\left(1+2\bar{\Psi}\left(\frac{\bar{\Psi}}{r}\right)'\times\right.\right.\nonumber \\
&&\left.\left.\left(1+\bar{\Psi}\left(\frac{\bar{\Psi}}{r}\right)'\right)\right)\right]\delta u+ 2 q \bar{\phi} \bar{\xi}^2 \delta w''+2q \bar{V}\bar{\phi}\left[ \sqrt{\bar{h}}\bar{\xi}'+\bar{h}\bar{\Psi} \left(\frac{\bar{\Psi}}{r}\right)'\left(\frac{1}{r}-\frac{\bar{V} \bar{\phi}' }{\bar{\phi}}+\frac{r }{b}-\frac{r }{b{\left(1-\frac{b\bar{\phi}'^2}{\bar{h}}\right)}^{\frac{1}{2}}}+\frac{3r}{L^{2}} \right)-\frac{m_{s}^{2}\bar{h}\bar{\Psi}^2}{r}\right.\nonumber \\
&&\left.\times\left(1+\bar{\Psi}\left(\frac{\bar{\Psi}}{r}\right)'\right)\right]\delta w' +2q \bar{\phi}\left[q^2 \bar{\phi}^2-\frac{ \bar{\xi} \bar{\xi}'}{r}+\bar{\xi} \bar{\Psi}' \left(\frac{\bar{\Psi}}{r}\right)'\left(\frac{\bar{\xi} \bar{\phi}'}{\bar{\phi}}-\bar{\xi}'-\frac{\bar{\xi}}{r }\right)+m_{s}^{2}\bar{V}\bar{h}\left(-1+\frac{\bar{\Psi}\bar{\Psi}'}{r}\right)\right]\delta w=0,
\end{eqnarray}
\begin{eqnarray}\label{30}
&&\delta \ddot{w}-\bar{\xi}^2\delta w''+\left[\frac{2q^2 \bar{\phi} \bar{\Psi}^2}{r^2  \bar{\phi}'}\left(\r \bar{\phi}' \bar{\phi}+ \bar{V}\bar{h}\right){\left(1-\frac{b\bar{\phi}'^2}{\bar{h}}\right)}^{\frac{3}{2}}-\bar{\xi} \bar{\xi}'\right] \delta w'-\left[\frac{2q^2 \bar{\phi} \bar{\Psi} \bar{\Psi}'}{r^2  \bar{\phi}'} \left(
r \bar{\phi} \bar{\phi}'+\bar{V}\bar{h}\right){\left(1-\frac{b\bar{\phi}'^2}{\bar{h}}\right)}^{\frac{3}{2}}+q^2 \bar{\phi}^2 \right.\nonumber \\
&&\left.-\frac{\bar{\xi} \bar{\xi}'}{r}-m_{s}^ 2 \bar{V}\bar{h}\right]\times \delta w -2q \bar{\phi}\left(1+\bar{\Psi} \left(\frac{\bar{\Psi}}{r}\right)'\right)\delta u-q \bar{\Psi} \delta \phi+\frac{q \bar{\phi} \bar{\Psi}}{\bar{\phi}'} \delta \phi'=0,
\end{eqnarray}
\begin{eqnarray}\label{31}
&&\frac{2q \bar{\Psi}}{r^2 \bar{\phi'}}\left(r\bar{\phi}'\bar{\phi}+\bar{V}\bar{h}{\left(1-\frac{b\bar{\phi}'^2}{\bar{h}}\right)}^{\frac{3}{2}}\right)\delta w''+ 2q \bar{\phi} \bar{\Psi} \left[\frac{\bar{\xi}'}{\bar{\xi} r}+\frac{3b\bar{V}\bar{h}}{2r^{2}\bar{\phi}'}{\left(1-\frac{b\bar{\phi}'^2}{\bar{h}}\right)}^{\frac{1}{2}}\times\left(\frac{-2b\bar{\phi}''\bar{\phi}'}{\bar{h}}+\frac{b\bar{h}'}{{\bar{h}}^{2}}\right)+\left(\frac{\sqrt{\bar{h}}\bar{\xi'}}{r^2 \bar{\phi} \bar{\phi}'}-\frac{2b\bar{V}\bar{\phi}'}{\bar{\phi}r^3}\right.\right.\nonumber\\
&&\left.\left.-\frac{2q^2 \bar{h} \bar{\Psi}^2}{r^4 \bar{\phi}'^2}{\left(1-\frac{b\bar{\phi}'^2}{\bar{h}}\right)}^{\frac{3}{2}}\right)\times{\left(1-\frac{b\bar{\phi}'^2}{\bar{h}}\right)}^{\frac{3}{2}}\right]\delta w'+\frac{2q  \bar{\phi} \bar{\Psi}}{r^2}\left[\frac{r q^2 \bar{\phi}^2}{\bar{\xi}^2}-\frac{\bar{\xi}'}{\bar{\xi}}-\frac{m_{s}^2 r}{\bar{V}}-\frac{3b\bar{V}\bar{h}\bar{\Psi}'}{2\bar{\phi}'\bar{\phi}\bar{\Psi}}{\left(1-\frac{b\bar{\phi}'^2}{\bar{h}}\right)}^{\frac{1}{2}}\times\left(\frac{-2b\bar{\phi}''\bar{\phi}'}{\bar{h}} \right.\right. \nonumber\\
&&\left.\left.+\frac{b\bar{h}'}{{\bar{h}}^{2}}\right)+\left(\frac{q^2 \bar{\phi}}{\bar{V} \bar{\phi}'}-\frac{m_{s}^2\bar{h}}{\bar{\phi}\bar{\phi}'}-\frac{\sqrt{\bar{h}}\bar{\xi}'}{r \bar{\phi} \bar{\phi}' }+\frac{2b\bar{V}\bar{\Psi}'\bar{\phi}'}{r\bar{\phi}\bar{\Psi}}+\frac{2q^2 \bar{h} \bar{\Psi}\bar{\Psi}'}{r^2 \bar{\phi}'^2}\times{\left(1-\frac{b\bar{\phi}'^2}{\bar{h}}\right)}^{\frac{3}{2}}\right)\times{\left(1-\frac{b\bar{\phi}'^2}{\bar{h}}\right)}^{\frac{3}{2}}\right]\delta w-2\left(\frac{\bar{\Psi}}{r}\right)'\delta u'-\nonumber\\
&&\left[ \left(\frac{\bar{\Psi}}{r}\right)'\left(\frac{1}{r}+\frac{\bar{\xi}'}{\bar{\xi}}\right)
+ \left(\frac{\bar{\Psi}}{r}\right)''-\frac{m_{s}^2\bar{\Psi}}{r \bar{V}}\right]\delta u+\left(\frac{\delta \phi'}{\bar{\phi}'} \right)'=0.
\end{eqnarray}
\subsection{Boundary condition and numerical results}
\begin{figure}[!ht]
\includegraphics[width=5.4cm,height=3.5cm]{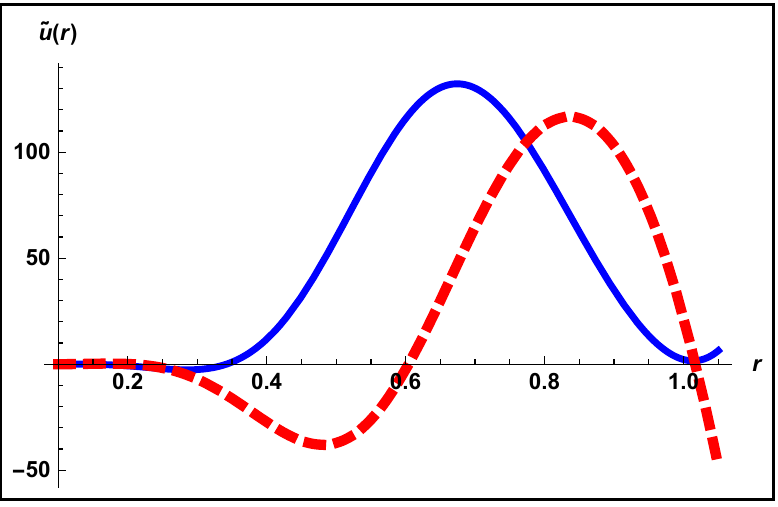}
\includegraphics[width=5.4cm,height=3.5cm]{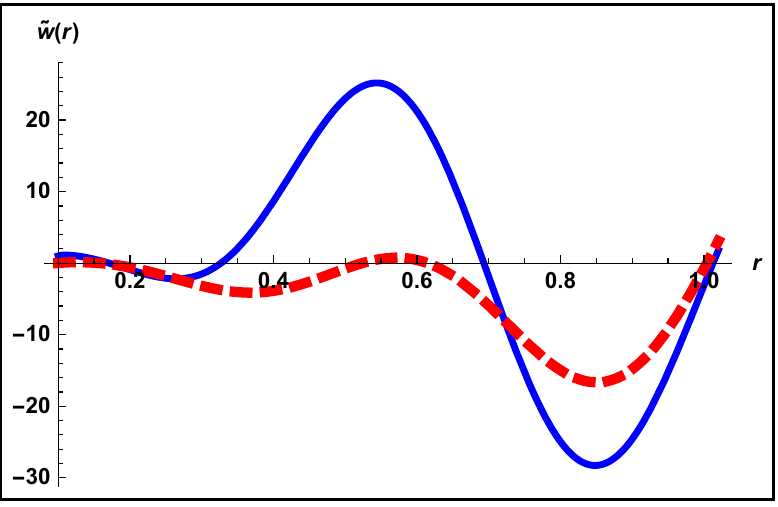}
\includegraphics[width=5.4cm,height=3.5cm]{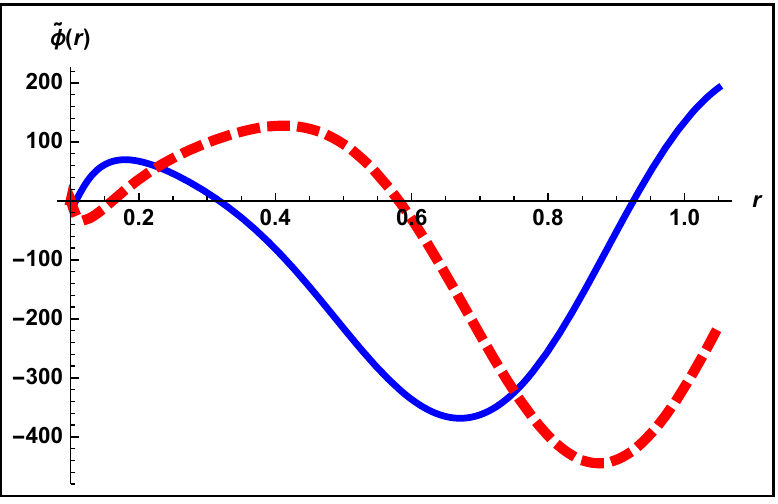}
 \caption{\footnotesize{Behavior of three perturbation functions $\tilde{u}(r)$, $\tilde{w}(r)$ and $\tilde{\phi}(r)$ for $q =20$, $\phi'_{+}=0.3$, $r_{+}=0.1$, $b=1$,  $\psi_{+}=0.2434$, with $\tilde{u}_{0}=0.0027+0.0009 i$ and $\omega=3.7 - 0.3715 i $. }}\label{fig9}
\end{figure}

To proceed further, we set the ingoing boundary condition near the event horizon for perturbation modes to $\delta u(t,r)=\mbox{Re}[ e^{-i\omega (t+r_{*})} \tilde{u}(r)]$ and make a Taylor expansion of the complex function $\tilde{u}(r)=\tilde{u}_{0}+\tilde{u}_{1}(r-r_{+})+\tilde{u}_{2} (r-r_{+})^2/2+...$ and for other perturbation modes in a similar fashion in equations (\ref{29}-\ref{31}), with the result
\begin{eqnarray}\label{305a}
&&\tilde{\phi}_{1}=\frac{-2q\psi_{+}{\omega}^{2}\left({\phi'_{+}}^{2}+\frac{V'_{+} h_{+}}{r_{+}}\left(1-\frac{b{\phi'_{+}}^2}{h_{+}}\right)^{\frac{3}{2}}\right)\tilde{w}_{0}+\left(\phi'_{+}V'_{+}\psi'_{+}h_{+}\left(\frac{2i\omega}{\sqrt{h_{+}}}-V'_{+}\right)+\phi'_{+}V'_{+}m_{s}^{2}\psi_{+}h_{+}\right)\tilde{u}_{0}}{\omega\left(\omega+iV'_{+}\sqrt{h_{+}}\right)},\nonumber\\
&&\tilde{w}_{1}=\frac{\left[\frac{V'_{+} \sqrt{h_{+}}}{r_{+}}+m_{s}^{2}\sqrt{h_{+}}-\frac{iq^2\psi_{+}^{2}\omega}{V'_{+}\sqrt{h_{+}}}\left(\frac{r_{+}{\phi'_{+}}^{2}}{V'_{+}\sqrt{h_{+}}}+\sqrt{h_{+}}\right)\times \left(1-\frac{b {\phi'_{+}}^{2}}{h_{+}}\right)^{\frac{3}{2}}\right]\tilde{w}_{0}-\frac{2q\phi'_{+}}{V'_{+}\sqrt{h_{+}}}\left(1+r_{+}\psi_{+}\psi'_{+}\right)\tilde{u}_{0}-\frac{i\omega q r_{+} \psi_{+}}{{V'_{+}}^{2}h_{+}}\tilde{\phi}_{1}}{V'_{+}\sqrt{h_{+}}-2i\omega},\nonumber \\
&&\tilde{u}_{1}=\frac{\left[\frac{V'_{+}\sqrt{h_{+}}}{r_{+}}-2\sqrt{h_{+}}{\psi'_{+}}^{2}r_{+}V'_{+}+m_{s}^{2}\sqrt{h_{+}}\left(1+2r_{+}\psi_{+}\psi'_{+}\right)\right]\tilde{u}_{0}-\left(\frac{2q\phi'_{+}{\omega}^{2}}{V'_{+}\sqrt{h_{+}}}+\frac{2i\omega q \phi'_{+}}{V'_{+}}\left(V'_{+}-m_{s}^{2}r_{+}\psi_{+}^{2}\right)\right)\tilde{w}_{0}}{V'_{+}\sqrt{h_{+}}-2i\omega} .
\end{eqnarray}
We fix $\tilde{w}_{0}=1$, integrating equations (\ref{29}-\ref{31}) using a shooting method with boundary conditions (\ref{305a}) where $\omega$ and $ \tilde{u}_{0}$ are shooting parameters, their value is determined in such a way that the perturbation modes $\tilde{u}(r)$ and $\tilde{w}(r)$ vanish at the reflective boundary.  Fig. \ref{fig9} and Tables \ref{tab1} and \ref{tab2}, give the shooting parameters at $T=T_{c}$ where the scalar field vanishes with the right choice of $\phi'_{+}$ and $\psi_{+}$. As can be seen,  $\mbox{Im}(\omega)$ is negative and consequently perturbation mode decay, rendering the hairy black hole stable.
\begin{table}
\centering
\setlength\tabcolsep{4pt}
\begin{minipage}{0.48\textwidth}
\centering
\caption{ The value of Shooting parameters for $q=20$, $r_{+}=0.1$, $\phi'_{+}=0.3$, $b=1$ and different values of $m_s^2$. }
\begin{tabular}{|c|c|c|c|}\hline
$m_{s}^{2}$ & $\psi_{+}$ & $\tilde{u}_{0}$ & $\omega$\\\hline
$0$ &$0.2434$ &$0.0027+0.0009 i$ &$3.7 - 0.3715 i $\\\hline
$0.5$ &$0.2396$ &$0.0027+0.00001 i$ &$3.70831 - 0.373831 i$\\\hline
$1$ & $0.235$&$0.003+0.005 i$ &$3.72278 - 0.375278 i$\\\hline
\end{tabular}
\label{tab1}
\end{minipage}%
\hfill
\begin{minipage}{0.48\textwidth}
\centering
 \caption{The value of Shooting parameters for $q=20$, $r_{+}=0.1$, $\phi'_{+}=0.3$, $m_{s}^{2}=0$ and different values of $b$.}
\begin{tabular}{|c|c|c|c|}\hline
$b$ & $\psi_{+}$ & $\tilde{u}_{0}$ & $\omega$\\\hline
$0.5$ &$0.232$ &$0.015+0.005 i$ &$3.66084 - 0.368084 i$\\\hline
$0.75$ &$0.238$ & $0.0035+0.0012 i$&$3.68 - 0.3695 i$\\\hline
$1$ &$0.2434$ &$0.0027+0.0009 i$ &$3.7 - 0.3715 i $\\\hline
\end{tabular}
 \label{tab2}
\end{minipage}
\end{table}
\section{Discussion and conclusions }\label{DC}
In this work we have shown that there are two types of linear instabilities in Einstein-Born-Infeld-scalar field theory; superradiant instability is related to small BI black holes and near horizon scalar condensation instability occurs for large BI black holes. The superradiant instability exits in a certain range of  frequencies and it is shown numerically that a larger BI coupling parameter leads to smaller $\mbox{Im}(\omega)$ and slower growth rate of the instability. Also, we demonstrated that there is no superradiant instability for large values of scalar mass. For large BI black holes the second instability is  tachyonic    when the effective mass violates the BF bound $m_{eff}^2<-\frac{9}{4 L^2}$.

These instabilities push a BI black hole towards a hairy black hole where the charged scalar hair is near the horizon. For superradiant instability, we showed that the solution space becomes larger by increasing the BI coupling parameter for a small hairy black hole which means that  there is more freedom to choose $\psi_+$, while the scalar charge and event horizon have an inverse relation to $\psi_+$. It was also shown that $T_c$ has an inverse relation to the event horizon and scalar mass and is related to the scalar charge according to $T_c \propto \sqrt{q}$,  where $T_c$ is the critical temperature and $\psi_2=0$. Interestingly, the metric solution, $h(r)$,
represents phase transition between a BI black hole and a hairy black hole, i.e. the small hairy black hole configuration is preferred for $T<T_c$. The system admits BI black hole configurations for $T>T_c$.  We numerically showed that as the BI coupling parameter increases, the critical temperature increases and phase transition becomes more probable in a small hairy black hole, in contrast to a planar hairy black hole where the critical temperature has an inverse relation to the BI coupling parameter. We predicted that the final state of superradiant instability is a small hairy black hole and showed numerically  that the small hairy black hole is stable at $T=T_{c}$ on account of $\mbox{Im}(\omega)<0$.
\vspace{3mm}
 \appendix
 \section{ Analytic calculation of quasinormal frequency}
 In what follows, we follow the procedure presented in \cite{her,li}. By definition $\Delta=V r^{2}$, $\Psi=\frac{\psi}{r}$ and $\lambda=l(l+1)$, equation (\ref{12} ) becomes
 \begin{equation}\label{eq30}
\Delta \frac{d}{dr}\left(\Delta \frac{d\Psi}{dr}\right)-\left(\lambda+m_{s}^{2}r^{2}\right) \Delta \Psi+r^4\left(\omega+qA_{t}\right)^{2}\Psi=0.
\end{equation}
To obtain the frequency analytically, we divide outside region of the horizon into two regions, near region, $r-r_{+}\ll \frac{1}{\omega}$ and far region, $r-r_{+}\gg r_{+}$. The matching method is then used in that we match solutions in an intermediate region and focus on small black holes $r_{+}\ll L$.
\subsection{Near region solution}
In this region, we can neglect scalar mass in the low frequency perturbation, $r\ll \frac{l}{m_{s}}$, since the scalar particle's Compton wavelength is much larger than the black hole horizon. Equation (\ref{eq30})
then became
\begin{equation}\label{eq31}
\Delta \frac{d}{dr}\left(\Delta \frac{d\Psi}{dr}\right)-\lambda \Delta \Psi+r_{+}^4\left(\omega+qA_{t}\right)^{2}\Psi=0.
\end{equation}
We now define a new variable $x=\frac{r-r_{+}}{r-r_{-}}$ and take $\Delta\equiv (r-r_{+})(r-r_{-})$, with the result
\begin{equation}\label{eq32}
x \frac{d}{dx}\left(x \frac{d\Psi}{dx}\right)+\left[{\bar{\omega}}^{2}-\frac{\lambda x}{(1-x)^{2}}\right]  \Psi=0.
\end{equation}
For convenience, we take $C=0$ in the equation (\ref{9}), where
\begin{equation}\label{eq33}
\bar{\omega}=\frac{r_{+}^{2}}{r_{+}-r_{-}}\left(\omega-\frac{qQ}{r_{+}}\times_{2}F_{1}
\left[\frac{1}{4},\frac{1}{2},\frac{5}{4},-\frac{Q^2b}{r_{+}^4}\right]\right).
\end{equation}
Superradiance regime now leads to $\omega<\frac{qQ}{r_{+}}\times_{2}F_{1}\left[\frac{1}{4},\frac{1}{2},\frac{5}{4},-\frac{Q^2b}{r_{+}^4}\right]$. One can get a solution of (\ref{eq32})
with ingoing wave boundary condition in terms of  hypergeometric functions
\begin{equation}\label{eq34}
\Psi\sim x^{-i \bar{\omega}}\left(1-x\right)^{\alpha}F\left(\alpha,\alpha-2i\bar{\omega},1-2i\bar{\omega};x\right),
\end{equation}
where $\alpha\equiv 1+l$. To match the far region solution, one expands the near region solution for large $r$. Now, using the properties of hypergeometric functions and the limit $x\longrightarrow 1$, we arrive at the following results
 \begin{eqnarray}
&&\Psi\sim \Gamma \left(1-2i \bar{\omega}\right)\left[\frac{\Psi_{\frac{1}{r}}^{near}}{r^{1+l}}+\Psi_{r}^{near}r^l\right],\label{eq35}\\
&&
\Psi_{\frac{1}{r}}^{near}\equiv \frac{\Gamma\left(1-2\alpha\right)\left(r_{+}-r_{-}\right)^{\alpha}}{\Gamma\left(1-\alpha\right)\Gamma\left(1-\alpha-2i \bar{\omega}\right)},\label{eq36}\\
&&
\Psi_{r}^{near}\equiv \frac{\Gamma\left(2\alpha-1\right)\left(r_{+}-r_{-}\right)^{1-\alpha}}{\Gamma\left(\alpha\right)\Gamma\left(\alpha-2i \bar{\omega}\right)}.\label{eq37}
\end{eqnarray}
Finally, this solution should be matched with that for a small $r$ limit.
\subsection{Far region solution}
In the far region, one takes $Q\longrightarrow 0$ and $M\longrightarrow 0$ to arrive at $\Delta=r^{2}\left(1+\frac{r^{2}}{L^{2}}\right)$. Then by defining a new variable $z\equiv 1+\frac{r^2}{L^{2}}$, equation (\ref{eq30})
became
\begin{equation}\label{eq38}
z\left(1-z\right)
\frac{d^{2}\Psi}{dz^{2}}+\left(1-\frac{5}{2}z\right)\frac{d\Psi}{dz}-\left[\frac{{\omega}^{2}L^2}{4z}+
\frac{\lambda}{4\left(1-z\right)}-\frac{m_{s}^{2}L^2}{4}\right]\Psi=0,
\end{equation}
with a solution as  hypergeometric function
\begin{equation}\label{eq38}
\Psi=z^{\frac{\omega L}{2}}\left(1-z\right)^{\frac{l}{2}}F\left(a,b,c;z\right),
\end{equation}
where
\begin{eqnarray}
&&a\equiv\frac{3}{4}+\frac{\omega L}{2}+\frac{l}{2}+\frac{1}{2}\sqrt{m_{s}^{2}L^2+\frac{9}{4}},\label{eq39} \\
&&b\equiv\frac{3}{4}+\frac{\omega L}{2}+\frac{l}{2}-\frac{1}{2}\sqrt{m_{s}^{2}L^2+\frac{9}{4}}.\label{eq41}\\
&&c\equiv 1+\omega L,\label{eq42}
\end{eqnarray}
Given the decaying boundary condition at infinity (\ref{13}), one obtains the far-region solution of equation (\ref{eq38}) as
 \begin{equation}\label{eq43}
\Psi\sim {(1-z)}^{\frac{l}{2}}z^{\frac{\omega L}{2}-a}F\left(a,1+a-c,1+a-b;\frac{1}{z}\right).
\end{equation}
To match the near-region solution, we study small $r$ behavior of the far-region solution by using $\frac{1}{z}\longrightarrow 1-z$ transformation properties of the hypergeometric function. For $z\longrightarrow1$ we have
\begin{equation}\label{eq44a}
F\left(a,1+a-c,1+a-b;\frac{1}{z}\right)\approxeq (z-1)^{c-a-b} \frac{\Gamma \left(1+a-b\right)\Gamma \left(a+b-c\right)}{\Gamma (a)\Gamma \left(1+a-c\right)}+\frac{\Gamma \left(1+a-b\right)\Gamma \left(c-a-b\right)}{\Gamma \left(1-b\right)\Gamma \left(c-b\right)}.
\end{equation}
The far-region solution (\ref{eq43}) in the limit of small $r$ becomes
\begin{eqnarray}
&&\Psi\sim \Gamma \left(1+a-b\right)\left[\frac{\Psi_{\frac{1}{r}}^{far}}{r^{1+l}}+\Psi_{r}^{far}r^{l}\right],\label{eq44}\\
&&
\Psi_{\frac{1}{r}}^{far}\equiv \frac{\Gamma\left(l+\frac{1}{2}\right)L^{1+l}}{\Gamma(a)\Gamma\left(1+a-c\right)},\label{eq45}\\
&&
\Psi_{r}^{far}\equiv \frac{\Gamma\left(-l-\frac{1}{2}\right)L^{-1}}{\Gamma(1-b)\Gamma\left(c-b\right)},\label{eq46}
\end{eqnarray}
where solution (\ref{eq44})
is for pure AdS. To have the regular condition  of the above solution at the origin we need
\begin{equation}\label{eq47a}
\Gamma\left(1+a-c\right)=\infty \Rightarrow 1+a-c=-N,
\end{equation}
for which we get a discrete spectrum of the frequency
\begin{equation}\label{eq47}
\omega_{N}L=2N+\frac{3}{2}+l+\sqrt{m_{s}^2L^2+\frac{9}{4}},
\end{equation}
where $N$ is a nonnegative integer number. However, the effects of the black hole add a correction to the AdS frequency as $\omega=\omega_{N}+i \delta$, $\delta$ which shows damping quasi-normal modes. For small black holes, we use approximation $\frac{1}{\Gamma(-N+\epsilon)}\sim (-1)^{N}N!\epsilon$, with the result
\begin{equation}\label{eq48}
\Psi_{\frac{1}{r}}^{far}=(-1)^{N+1}i \delta N! \frac{\Gamma\left(l+\frac{1}{2}\right)L^{2+l}}{2\Gamma(a)}.
\end{equation}
\subsection{Matching solutions}
At this stage, we match near and far region solutions in an intermediate region with matching condition $\Psi_{r}^{near} \Psi_{\frac{1}{r}}^{far}=\Psi_{r}^{far}\Psi_{\frac{1}{r}}^{near}$. Then
\begin{equation}\label{eq49}
\delta=2i\frac{{(r_{+}-r_{-})}^{2\alpha-1}}{L^{2l+2} } \frac{\Gamma\left(1-2\alpha\right)\Gamma(\alpha)}{\Gamma\left(2\alpha-1\right)\Gamma\left(1-\alpha\right)}\times \frac{(-1)^{N}}{N!} \frac{\Gamma(a)}{\Gamma\left(1-b\right)\Gamma\left(c-b\right)}\frac{\Gamma\left(-l-\frac{1}{2}\right)
\Gamma\left(\alpha-2i\bar{\omega}\right)}{\Gamma\left(l+\frac{1}{2}\right)
\Gamma\left(1-\alpha-2i\bar{\omega}\right)},
\end{equation}
The properties  of Gamma function \cite{her,abra} for nonnegative $l$ are given by
\begin{equation}\label{eq50}
\frac{\Gamma(a)}{\Gamma\left(1-b\right)\Gamma\left(c-b\right)}\frac{\Gamma
\left(-l-\frac{1}{2}\right)}{\Gamma\left(l+\frac{1}{2}\right)}=\frac{(-1)^{N}
\Gamma\left(N+\frac{3}{2}+l+\sqrt{m_{s}^{2}L^{2}+\frac{9}{4}}\right)}
{\Gamma\left(N+l+\sqrt{m_{s}^{2}L^{2}+\frac{9}{4}}\right)
\Gamma\left(l+\frac{1}{2}\right)}\times\prod_{k=1}^{N}\left(l+\frac{1}{2}+k\right) ,
\end{equation}
\begin{eqnarray}
&&\frac{\Gamma\left(1-2\alpha\right)\Gamma\left(\alpha\right)}
{\Gamma\left(2\alpha-1\right)\Gamma\left(1-\alpha\right)}=
\frac{(-1)^{l+1}}{2}\frac{(l!)^{2}}{(2l)!(2l+1)!} ,\label{eq51}\\
&&
\frac{\Gamma\left(\alpha-2i\bar{\omega}\right)}
{\Gamma\left(1-\alpha-2i\bar{\omega}\right)}=
(-1)^{l}2i\bar{\omega}\prod_{k'=1}^{l}\left({k'}^{2}+4{\bar{\omega}}^{2}\right).\label{eq52}
\end{eqnarray}
One now  gets
\begin{equation}\label{eq53}
\delta=-2 \bar{\omega}\frac{{(r_{+}-r_{-})}^{1+2l}}{N!L^{2l+2}}\frac{(l!)^{2}}{(2l)!\left(2l+1\right)!}
\times\frac{\Gamma\left(N+\frac{3}{2}+l+\sqrt{m_{s}^{2}L^{2}+\frac{9}{4}}\right)}{\Gamma\left(N+l+
\sqrt{m_{s}^{2}L^{2}+\frac{9}{4}}\right)\Gamma\left(l+\frac{1}{2}\right)} \times\prod_{k=1}^{N}\left(l+\frac{1}{2}+k\right)\prod_{k'=1}^{l}\left({k'}^{2}+4{\bar{\omega}}^{2}\right).
\end{equation}
Due to the AdS/CFT correspondence we need to have $C\neq 0$, which just shifts $\mbox{Re}(\omega)$ as $\mbox{Re}(\omega)\longrightarrow \mbox{Re}(\omega)+q C$ without changing $\mbox{Im}(\omega)$ \cite{her}.

\end{document}